\documentclass[twocolumn]{aastex62}
\pdfoutput=1 
\usepackage{appendix}
\usepackage{amsmath,amstext}
\usepackage[T1]{fontenc}
\usepackage[figure,figure*]{hypcap}
\graphicspath{{./}{figures/}}

\received{?}
\revised{?}
\accepted{?}
\submitjournal{AAS}

\shorttitle{Revisitng the MMEN with precise stellar parameters}
\shortauthors{Dai et al.}

\begin{document}

\title{CKS IX: Revisiting the Minimum-Mass Extrasolar Nebula with Precise Stellar Parameters}

\author{Fei Dai}
\affiliation{Division of Geological and Planetary Sciences
1200 E California Blvd, Pasadena, CA, 91125, USA}
\affiliation{Department of Physics and Kavli Institute for Astrophysics and Space Research, Massachusetts Institute of Technology, Cambridge, MA, 02139, USA}
\affiliation{Department of Astrophysical Sciences, Princeton University, 4 Ivy Lane, Princeton, NJ, 08544, USA}
\email{fdai@caltech.edu}

\author{Joshua N.\ Winn}
\affiliation{Department of Astrophysical Sciences, Princeton University, 4 Ivy Lane, Princeton, NJ, 08544, USA}

\author{Kevin Schlaufman}
\affiliation{Department of Physics and Astronomy, Johns Hopkins University, 3701 San Martin Drive, Bloomberg 366,Baltimore, MD, 21218, USA}

\author{Songhu Wang}
\affiliation{Department of Astronomy, Yale University, New Haven, CT 06511, USA}

\author{Lauren Weiss}
\affiliation{Institute for Astronomy, University of Hawaii at Manoa, Honolulu, HI 96822, USA}

\author{Erik A. Petigura}
\affiliation{Department of Physics \& Astronomy, University of California Los Angeles, Los Angeles, CA 90095, USA}

\author{Andrew W.\ Howard}
\affiliation{Cahill Center for Astronomy and Astrophysics,
1200 E California Blvd, Pasadena, CA, 91125, USA}

\author{Min Fang}
\affiliation{Cahill Center for Astronomy and Astrophysics,
1200 E California Blvd, Pasadena, CA, 91125, USA}



\begin{abstract}
 \noindent Previous works on Kepler multi-planet systems revealed a remarkable intra-system uniformity in planet radius/mass; moreover the average planet size increases with host mass/metallicity. This observation provides tantalizing evidence that the outcome of planet formation can be linked to the
properties of the host star and its disk. In a simple in-situ formation scenario, the minimum-mass
extrasolar nebula (MMEN) reconstructed from a planetary system reflect the properties of its natal
disk.Specifically, one might expect a one-to-one correspondence between the solid surface density of
the MMEN and the host star mass $M_\star$ and metallicity [Fe/H]. Leveraging on the precise host star properties from the California-{\it Kepler}-Survey (CKS), we found that $\Sigma=$ $50^{+33}_{-20} \rm{~g~cm}^{-2}$ $(a/{\rm AU})^{-1.75\pm0.07}$ $(M_\star/M_\odot)^{1.04\pm0.22}$ $10^{0.22\pm0.05{\rm [Fe/H]}}$ for {\it Kepler}-like systems (1-4$R_\oplus$; $a<$1AU). The strong $M_\star$ dependence is reminiscent of previous dust continuum results that the solid disk mass scales with $M_\star$. The weaker [Fe/H] dependence shows that sub-Neptune planets, unlike giant planets, form readily in lower-metallicity environment. The innermost region ($a<$ 0.1AU) of a MMEN maintains a smooth profile despite a steep decline of planet occurrence rate: a result that favors the truncation of disks by co-rotating magnetospheres with a range of rotation periods, rather than the sublimation of dusts. The $\Sigma$ of {\it Kepler} multi-transiting systems shows a much stronger correlation with $M_\star$ and [Fe/H] than singles. This suggests that the dynamically hot evolution that produced single systems also partially removed the memory of formation in disks. Radial-velocity planets yielded a MMEN very similar to CKS planets; whereas transit-timing-variation planets' postulated convergent migration history is supported by their poorly constrained MMEN. We found that lower-mass stars have a higher efficiency of forming/retaining planets:  for sun-like stars about 20\% of the solid mass within $\sim$1AU are converted/preserved as sub-Neptunes, compared to 70\% for late-K-early-M stars. This may be due to the lower binary fraction, lower giant-planet occurrence or the longer disk lifetime of lower-mass stars.

\end{abstract}

\keywords{planets and satellites: formation}


\section{Introduction} \label{sec:intro}
The protoplanetary disk sets the stage for planet formation. Disk properties may play a decisive role in planet formation by changing the availability of solid materials, the rate of core assembly and gas accretion and the overall time-span allowed for planet formation. The occurrence of giant planets correlates strongly with the mass and metallicity of host stars \citep[e.g.][]{Santos2004,Fischer2005,Johnson2010}. The metallicity correlation is often quoted as a ''smoking gun" for the core accretion theory \citep[e.g.][]{Safronov,POLLACK199662} in which the efficiency of planet formation is strongly dependent on the density of solid materials in a disk. However, whether such a strong correlation extends to smaller planets is unclear \citep[e.g.][]{Wang2015,Buchhave,Schlaufman2015, CKS4}. \citet{Buchhave} reported a null result when they compared the average metallicity of {\it Kepler} planet hosts with a control sample of field stars with no transiting planets. As pointed out by \citet{Zhu2016}, a large fraction of this control sample are in fact stars hosting undetected planets due to the high occurrence rate of sub-Neptune planets \citep[e.g.][]{Fressin} and the low transit probability. Any statistical comparison with such a contaminated control sample will have reduced significance. Another approach to study the occurrence-metallicity correlation directly compares the average number of planets per star for different stellar metallicity bins \citep{Wang2015,CKS4}. \citet{CKS4} reported that the occurrence for sub-Neptune planets with orbital periods 1-10 days is significantly higher in metal-rich ([Fe/H]$>$0) systems, whereas this preference for metal-rich systems diminishes for longer orbital periods (10-300 days). 

\citet{Weiss}, \citet{Millholland} and \citet{Songhu} revealed a remarkable intra-system uniformity of multi-planet systems: sibling planets are similar both in mass and radius. Furthermore, the average planet radii/masses in a particular system are correlated with host star mass and metallicity \citep{Millholland}. These observations hint that the outcome of planet formation is at least partially determined by the properties of the protoplanetary disk and the host star. One may speculate that a more massive, metal-rich host star is accompanied by a more massive, solid-enhanced protoplanetary disk that, in turn, form more massive planetary cores. The continuum observations in millimeter wavelength by \citet{Andrews} already established a linear correlation between stellar mass and disk dust mass  (integrated over a radial extent of tens to hundreds of AU, while also showing a substantial scatter of 0.7 dex). Another intriguing result from \citet{Weiss} is that when planets are smaller in size, they also have smaller orbital spacing. This observed planet radii-orbital spacing correlation may  stem from the variation of the initial separation of planet embryos: those that happen to be closer to each other will compete for solid materials in the disk and end up being closely-packed smaller planets.

 The minimum-mass solar nebula \cite[MMSN,][]{Weidenschilling,Hayashi} puts a lower limit on density profile of the protoplanetary disk in which the Solar system planets formed.  The construction of the MMSN spreads out the masses of the planets at their current orbits into an area determined by their orbital spacings. This process effectively assumes the {\it in-situ} formation of the planets and the planets only accreted material from their local feeding zones. The classic study of \citet{Hayashi} found that the total surface density (gas and dust) can be described by $\Sigma_{\rm tot}$ $\approx$ 1700 $(a/{\rm AU})^{-1.5}$g cm$^{-2}$ or $\Sigma$ $\approx$ 7.1 $(a/{\rm AU})^{-1.5}$g cm$^{-2}$ for the dust component only. \citet{CL2013} used the transiting exoplanets discovered by the {\it Kepler} mission to construct the minimum-mass extrasolar nebula (MMEN). They found that the average MMEN disk profile is similar to but denser than the MMSN: $\Sigma$ $\approx$ 50 $(a/{\rm AU})^{-1.6}$g cm$^{-2}$. They also performed a series of order-of-magnitude calculations to show that the {\it in-situ} formation of {\it Kepler}-like planets should be fast and efficient in such a disk. 
 
 At the time of writing \citet{CL2013}, little was known about the properties of the {\it Kepler} stars, precluding an investigation of MMEN as a function of host star properties. Thanks to the California-{\it Kepler}-Survey \citep[CKS;][]{CKS1}, the high-resolution spectra of $\sim1300$ {\it Kepler} planet host stars have been obtained. Combining the spectroscopic and parallax information from {\it Gaia} \citep{Gaia}, the CKS team derived precise stellar and planetary parameters of {\it Kepler} planets \citep{Fulton2018}. In this work, we revisit the problem of MMEN, specifically we ask: is there a correlation between the solid surface density of a MMEN $\Sigma$ and stellar mass $M_\star$ and metallicity [Fe/H]? If so, what is the implication for the formation of {\it Kepler}-like planetary systems? This MMEN approach is complementary to the traditional occurrence rate study \citep[e.g.][]{CKS4}, as MMEN simultaneously brings the occurrence rate, the planet multiplicity, the orbital spacing and the size of the planets into perspective. We hope this will shed new light on the formation of sub-Neptune planets.

This paper is organized as follows. In Section 2, we define the samples of planetary systems used in this work. We present the construction and modeling of the MMEN in Section 3. In Section 4, we discuss the results and the implications for the formation of sub-Neptune planets. We conclude the findings of this work in Section 5.

\section{The samples}\label{sec:sample}

\subsection{The CKS Sample}
Our study is mostly based on the California-{\it Kepler}-Survey \citep{CKS1} given its uniformity and precision. We started with the whole CKS catalog which consists of the 960 bright ({\it Kepler} Magnitude $<$14.2) planet hosts, the multi-planet host stars (484), hosts of ultra-short period planets (USP: $P_{\rm orb}$< 1 day; 127) and 109 host stars of various other types of planets. We incorporated the new stellar/planetary properties from \citet{Fulton2018} that included {\it Gaia} parallax constraints \citep{Gaia} and dilution factors from neighboring stars. We imposed several filters to identify a more uniform subsample:

{\begin{itemize}
    \item We restricted our attention to the brightest host stars ({\it Kepler} Magnitude $<$14.2).
    \item We removed any system that is designated as a False Positive by the {\it Kepler} team.

 \item Since planets around evolved stars are harder to detect and usually have larger radius uncertainty, we excluded any evolved star with the {\it ad hoc} relation suggested by \citet{CKS3}:
\begin{equation}
R_\star/R_\odot> 10^{0.00025(T_{\rm eff}/{\rm K}-5500)+0.20}
\end{equation}

\item We only included host stars with $T_{\rm eff}$ between 4700K and 6500K to avoid the known limitations of synthetic stellar atmospheric models \citep{CKS1},

\item We focused on the sub-Neptune planets ($1R_\oplus<R_p<4R_\oplus$) in this work. Since most of the {\it Kepler } planets do not have mass constraints, we have to rely on the measured radii and a mass-radius relationship to infer the planetary masses. Previous works \citep[e.g.][]{WolfgangLopez2015} showed that planets in the sub-Neptune radius range likely has a rocky core and a H/He envelope of $\lesssim 10$\% in mass. In other words, their mass is dominated by the mass of the core. Therefore the bulk mass of the sub-Neptune planets roughly corresponds to the amount of solid disk materials in them. In contrast, the mass of giant planets is dominated by the H/He gas. There is a large uncertainty in the core mass of giant planets; even the core mass of Jupiter is uncertain \citep{Wahl2017}. The cut at $4R_\oplus$ also excludes the emerging class of ``super-puff'' planets where mass-radius relationships fail because the transit radii are likely inflated by high-altitude dusts or aerosols \citep{Roberts,WangDai,Gao}.

\item We excluded the ultra-short-period planets because most of them were discovered by an independent study \citep{USP} with a Fourier-based transit detection method. On the other hand, most {\it Kepler} planets were identified by the {\it Kepler} team with a wavelet-based matched filter \citep{Jenkins}. The USPs likely have different detection probability from the other {\it Kepler} planets. Moreover, the larger mutual inclination and orbital period spacing of USP planets both suggest orbital migration \citep{Dai}, thus making it difficult to incorporate the USPs in the fundamentally {\it in-situ} MMEN framework.

\item Planets with grazing transits (impact parameter b > 0.9) were excluded. A grazing orbit often produces V-shaped transits. V-shaped transits suffer from a higher false positive probability. In addition, transit parameters tend to be degenerate with each other, leading a larger uncertainty in the planetary radius.
\end{itemize}}

After the various cuts described above, we are left with a sample size of 1041 planets around 712 stars. Fig \ref{the_sample} shows the distribution of the effective temperature $T_{\rm eff}$, metallicity [Fe/H] of the host stars, as well as the measured radii $R_{\rm p}$, inferred masses $M_{\rm p}$ and orbital periods $P_{\rm orb}$ of the planets.

\subsection{The KOI, RV and TTV Sample}
The CKS stellar properties are most precise for stars with $4700K<T_{\rm eff}<6500K$ where synthetic spectral models perform best \citep{CKS1}. We therefore used the Kepler Object of Interest (KOI) planets \citep{Mathur} for a sample of broader dynamical range in spectral types, $M_\star$ and [Fe/H]. We note that the stellar properties in the KOI sample are less precise than the CKS sample and are derived from a hodgepodge of methods including asteroseismology, photometric bands, spectroscopy etc. Therefore the conclusions of this paper is based on the analysis of the CKS sample, the KOI sample merely served as a consistency check. We applied the same sample selections as in the previous section except that we removed the cut on $T_{\rm eff}$. This resulted in a total of 1197 planets around 851 stars for KOI sample.

For both the CKS and the KOI sample, we had to rely on mass-radius relationships to estimate the masses of the planets from the transit radii. For more definite mass constraints, we used planets with transit-timing variations (TTV) or radial-velocity (RV) mass constraints. We used the TTV sample of 145 planets in 55 systems from the uniform analysis of \citet{Hadden}. We adopted the CKS stellar parameters for these systems. For the RV sample, we queried the {\sc Exoplanet Archive}\footnote{\url{https://exoplanetarchive.ipac.caltech.edu/}} for the stellar and planetary parameters of planets with a $M_{\rm p}$sin$i<30M_\oplus$ and at least 3-$\Sigma$ confidence. The query resulted in 135 planets from 51 systems.

\section{Constructing the MMEN}
\subsection{Planet Radius to Mass}
Assuming that close-in sub-Neptune planets formed {\it in-situ}, we can use each planet to sample the local solid surface density of its disk. For the CKS and KOI sample, we start by converting the observed transit radii to planetary masses using various mass-radius relationships reported in the literature.

1) \citet{Lissauer2011} proposed a simple power law based on the six solar system planets in the mass range between Mars and Saturn:
\begin{equation}
\frac{M_{\rm p}}{M_\oplus} =(\frac{R_{\rm p}}{R_\oplus})^{2.06}
\end{equation}

2) \citet{Weiss2014} presented a sample of 65 sub-Neptunes with orbital period <100 days. The masses of these planets were derived from both radial velocity follow-up and transit timing variations. \citet{Weiss2014} proposed a separate mass-radius relationship for superEarths (R$_p$<1.5$R_\oplus$) and mini-Neptunes (1.5$R_\oplus$<R$_p$<4$R_\oplus$):
\begin{equation}
R_{\rm p}<1.5R_\oplus: \rho_{\rm p} = 2.43+3.39(\frac{R_{\rm p}}{R_\oplus}) \rm{~g~cm}^{-3}
\end{equation}
\begin{equation}
1.5R_\oplus<R_{\rm p}<4R_\oplus: \frac{M_{\rm p}}{M_\oplus} = 2.69(\frac{R_{\rm p}}{R_\oplus})^{0.93}
\end{equation}

3) \citet{Wolfgang2016} calibrated a mass-radius relationship using $\sim 80$ sub-Neptune planets (<4$R_\oplus$) from RV mass measurements. The analysis was done in a Hierarchical Bayesian framework. 

\begin{equation}\label{MR}
M_{\rm p}/M_\oplus = 2.7 (R_{\rm p}/R_\oplus)^{1.3}
\end{equation}
They also reported a 1.9 $M_\oplus$ Gaussian dispersion as the intrinsic scatter in the mass distribution.

4) The mass-radius relationship proposed by \citet{Chen2017} is designed to be applicable to asteroid-sized objects all the way to stars. Calibrated with more than 300 solar-system and extra-solar objects, the mass-radius relationship span more than eight orders of magnitude in mass. Specifically, in the sub-Neptune regime, \citet{Chen2017} drew a distinction between planets smaller or larger than 2.0$^{+0.7}_{-0.6} M_\oplus$:
\begin{equation}
M_{\rm p}<2M_\oplus: R_{\rm p} \sim M_{\rm p}^{0.28} 
\end{equation}
\begin{equation}
M_{\rm p}>2M_\oplus: R_{\rm p} \sim M_{\rm p}^{0.59} 
\end{equation}

We used the {\sc Forecaster}\footnote{\url{https://github.com/chenjj2/forecaster}} code provided by \citet{Chen2017} to sample the masses of the CKS planets from their measured radii. 

5) More recently, using an updated list of both radial velocity mass measurements and transit timing variations, \citet{Mills} proposed the following power law relation:

\begin{equation}
\frac{M_{\rm p}}{M_\oplus}  = 10^{0.45} (\frac{R_{\rm p}}{R_\oplus})^{0.99}
\end{equation}

 We employed the mass-radius relationship from \citet{Wolfgang2016} for the purpose of generating figures, but our quantitative results account for the systematic uncertainties between these mass-radius relationships. For the TTV and RV samples, we skipped this step and used the reported mass constraints directly.

\subsection{Mass to Solid Surface Density}\label{feeding_zone}
Given that sub-Neptunes usually have $\lesssim 10$\% of their mass in H/He envelopes \citep{WolfgangLopez2015}, the bulk mass of these planets is also approximately the mass of solid materials they contain.  We remind the readers that the MMEN analysis implicitly assumes that the planets themselves formed and remained near their current-day orbits; and their only accreted solid materials from its neighborhood: the feeding zone. We then calculated a solid surface density $\Sigma$ assuming different widths of local feeding zones:

1) The first prescription assumes the planet accreted solids from a feeding zone whose width is proportional to the planet's Hill radius:
\begin{equation}
\label{hill}
\Sigma= \frac{M_{\rm p}}{2\pi a \Delta a}
\end{equation}

\begin{equation}\label{a}
a = (\frac{GM_\star P_{\rm orb}^2}{4\pi^2})^{1/3}
\end{equation}

\begin{equation}
\Delta a = k R_{\rm Hill}
\label{eqn:delta_a}
\end{equation}

\begin{equation}
R_{\rm Hill} \equiv a(\frac{M_{\rm p}}{3M_\star})^{1/3}
\end{equation}

\noindent where $\Sigma$ is the solid surface density; $a$ is the semi-major axis of the planet; and k is a constant. We chose k = 10 because \citet{Weiss} found empirically that typical orbital spacing between {\it Kepler} multi-planet systems is 10 mutual Hill radii or larger.

2) \citet{Schlichting2014} argued that planets can have a much larger effective feeding zone if we account for giant impact collisions. This entails swapping Eqn \ref{eqn:delta_a} with:

\begin{equation}
\Delta a = 2^{3/2}a(\frac{a}{R_p}\frac{M_p}{M_\star})^{1/2}
\end{equation}

3) We also used the simple prescription of \citet{CL2013} for easier comparison with their results:

\begin{equation}
\label{sigma}
\Sigma= \frac{M_{\rm p}}{2\pi a^2}
\end{equation}

In this prescription, each planet is spread out into an annulus of a width equal to its semi-major axis.

4) Finally we tried the prescription of \citet{Raymond}. Here neighboring planets' feeding zones are separated by the geometric means of their semi-major axes. For the innermost and outermost planet the feeding zone extend to $a_{\rm in}/1.5$ and $1.5a_{\rm out}$.  We found that this prescription is more susceptible to missing (non-transiting) planets. With a missing planet, the neighboring planets have much larger feeding zones and in turn give rise to anomalously low $\Sigma$. We will return to discuss this point further in Section \ref{sec:universal}.

We found that these prescriptions do not change the qualitative conclusions of this work, it generally only affects overall normalization of $\Sigma$. In the subsequent part of the paper, we will stick to the first prescription for its simplicity and its robustness against missing planets. In the MMSN work \citep{Hayashi}, the solid surface density $\Sigma$ was augmented by a gaseous component such that the resultant protoplanetary disk has solar abundance. However, recent {\it ALMA} observations \citep{Ansdell} showed that the gas-to-dust ratio in a protoplanetary disk, particularly in the inner disk, deviates significantly from the 100-200 assumed for the interstellar medium (ISM). Therefore, we focused on the solid surface density $\Sigma$ without augmenting it with a gaseous component. 

\subsection{Transit Probability and Detection Bias}
The detection of transiting planets suffers from two major biases: transit probability and detection bias. If unaccounted for, these two effects result in a bias towards shorter-period, larger planets. The transit probability is a geometric effect. In order for an observer to see a transit, the inclination of a planet's orbit must be close to 90$^\circ$. In the limit of small planets ($R_{\rm p}<<R_\star$) and low eccentricity, the transit probability is given by:
\begin{equation}
p_{\rm tra} = 0.9\frac{R_\star}{a}
\end{equation}
Note that we have included a factor of 0.9 since we only considered planets with impact parameter b$<$0.9.

The detection bias quantifies how complete a particular transit search pipeline is in detecting {\it bona-fide} planets. For {\it Kepler}, given the complexity of the entire pipeline, the completeness of the pipeline was determined empirically with injection-recovery tests. Previous works \citep{Christiansen,Petigura2013} have done extensive injection-recovery tests to characterize the behavior of the {\it Kepler} pipeline. Here, we adopt \citet{CKS3}'s  formulae for the detection probability as a function of signal-to-noise ratio $p_{\rm det} ({\rm SNR})$ (see their Eqn 2 and 3). 

To debias the CKS, KOI and TTV sample, we simply combined the transit probability and the pipeline detection probability. In our subsequent analyses, each planet in our sample was given a statistical weight inversely proportional to the combined probability:
\begin{equation}\label{weight}
w = \frac{1}{p_{\rm tra}p_{\rm det}}
\end{equation}

For the RV sample, we caution the readers that we could not find a simple way to correct for detection bias given the inhomogeneity of RV surveys. We chose to leave it uncorrected and assigned statistical weight only based on measurement uncertainties. In Section \ref{rv-ttv}, we will see that the $\Sigma$ inferred from RV sample is very similar but slightly denser than that of the transiting planets. This may relate to a detection or at least a publication bias towards heavier planets in RV surveys.

\subsection{Power Law Model}\label{sec:powerlaw}
\begin{figure*}
\begin{center}
\includegraphics[width = 1.9\columnwidth]{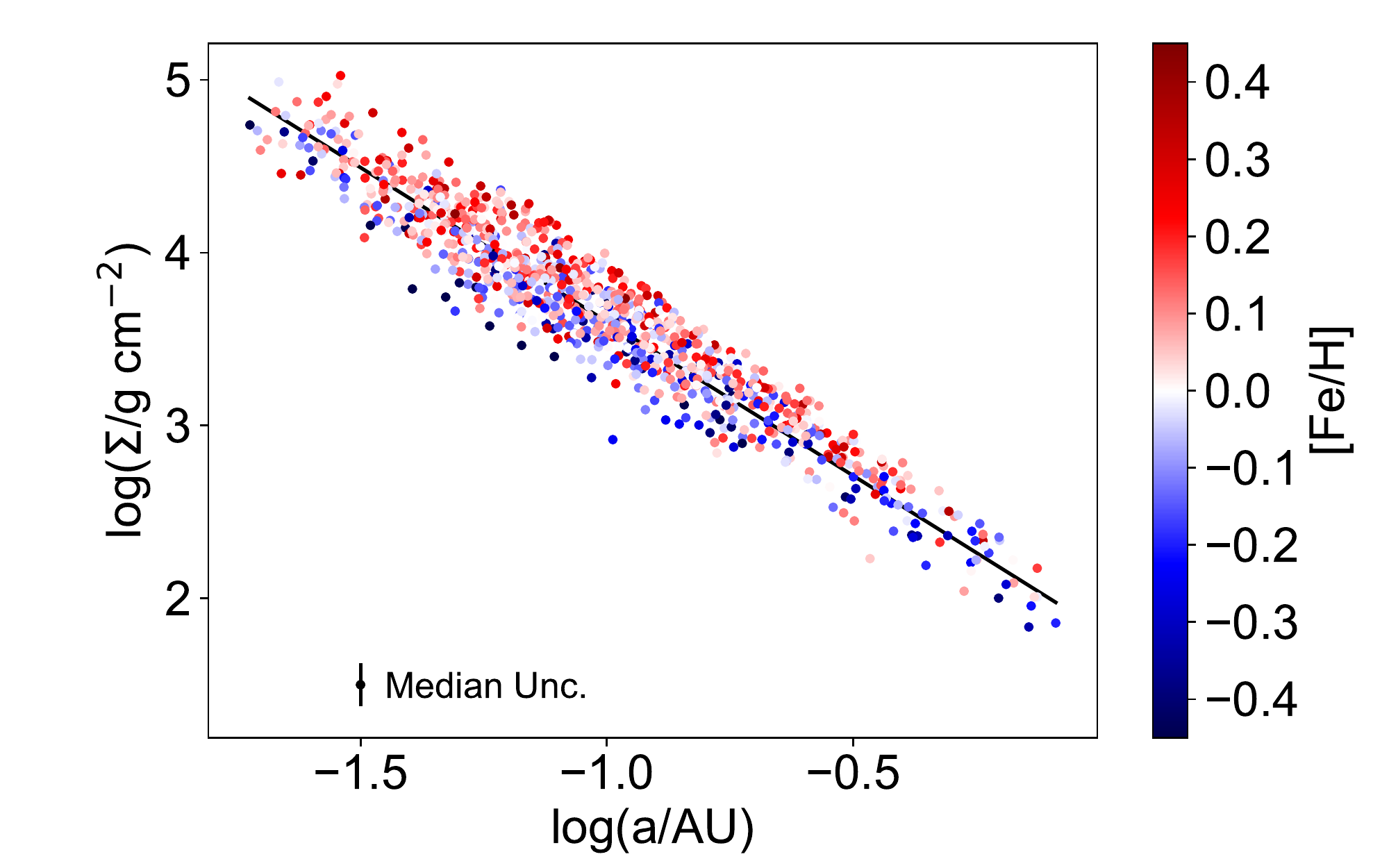}
\includegraphics[width = 1.9\columnwidth]{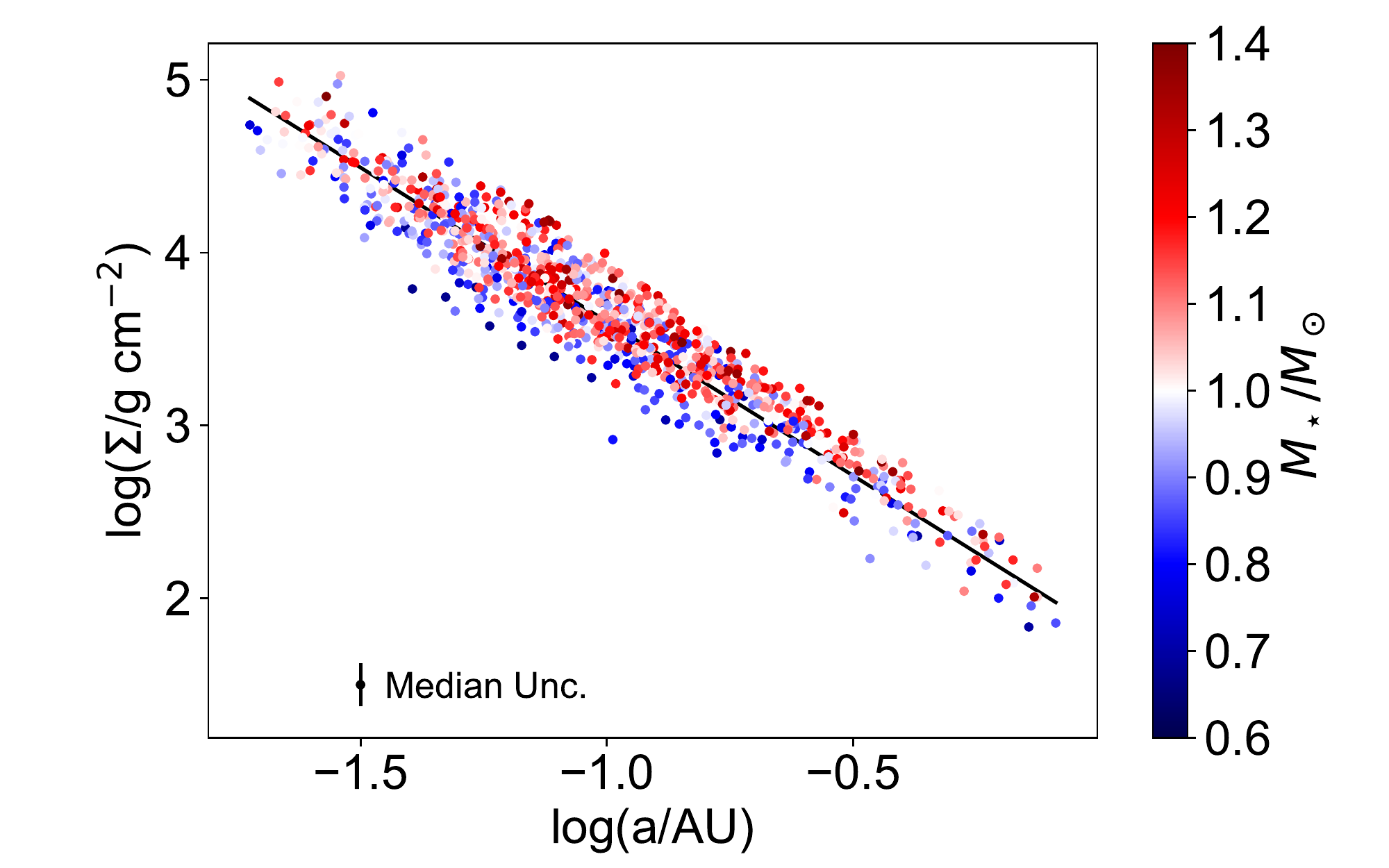}
\caption{The solid surface densities $\Sigma$ as inferred from the CKS planets against the semi-major axes. Here, we used the mass-radius relationship of \citet{Wolfgang2016} and assumed a feeding zone width of 10 Hill Radii. The surface density is well described by a simple power law relation with the orbital distance: $\Sigma\propto$$(\frac{a}{\rm AU})^{-1.7}$ (black line). In the two panels, the colorbars respectively encode the stellar metallcity [Fe/H] and the stellar mass $M_\star$ both of which show a positive correlation with the residual variation on $\Sigma$.  We caution the reader of a correlation between $M_\star$ and [Fe/H] in the CKS sample \citep{CKS4}. We performed a series of quantitative tests to minimize this correlation, see Section \ref{sec:strongMs}.}
\label{a_vs_sigma}
\end{center}
\end{figure*}

We plot the solid surface density $\Sigma$ reconstructed from the CKS planets as a function of orbital distance $a$ in Fig. \ref{a_vs_sigma}. Here, we employed the mass-radius relationship from \citet{Wolfgang2016} and assumed feeding zone widths given by prescription (1) in Section \ref{feeding_zone}. The solid surface density follows a tight relation with the semi-major axis $a$ similar to the power law profile previously reported for MMSN \citet{Hayashi} and MMEN \citep{CL2013}. This tight correlation with $a$ partially stems from the construction of MMEN itself. If all planets had equal masses, the solid surface density would be $\Sigma \propto a^{-2}$ with the (1) and (3) prescription of feeding zone width in Section \ref{feeding_zone} or $\Sigma \propto a^{-2.5}$ with the (2) prescription. 

We colorcode the planets with the host star mass $M_\star$ and metallicity [Fe/H]. Visually, both $M_\star$ and [Fe/H] are correlated with the residual variation in $\Sigma$. This trend is more obvious if we divide the sample into quartiles of $M_\star$ and [Fe/H] in Fig. \ref{four_panel}. The metal-rich systems and the more massive systems tend to have a larger $\Sigma$. To quantify the above statement, we fit the correlations between $\Sigma$, $a$, $M_\star$ and [Fe/H] with a simple power law model in the form:

\begin{equation}
\Sigma= \Sigma_0~(\frac{a}{\rm AU})^l~10^{m {\rm[Fe/H]}}~(\frac{M_\star}{M_\odot})^n
\end{equation}

\noindent where $\Sigma$ is the solid surface density; $\Sigma_0$ is a normalization constant; $a$ is the orbital distance; [Fe/H] is the host star metallicity proxied by the iron content and $M_\star$ is the mass of the host star. The equation is easier to work with in logarithmic space:
\begin{equation}
\label{eqn:powerlaw}
\rm{log}(\Sigma) = \rm{log}(\Sigma_0)+ l~\rm{log}(a/\rm{AU})+m~\rm{[Fe/H]}+n~\rm{log}(M_\star/M_\odot)
\end{equation}
the set of free parameters in this model are log($\Sigma_0$), $l$, $m$ and $n$. We also included a Gaussian dispersion $\Delta$log($\Sigma_0$) around log($\Sigma_0$) to account for any intrinsic scatter in the solid surface density in addition to the measurement uncertainties.

We note that when the independent variable has measurement uncertainty, the ordinary least-square estimator (OLS) tend to underestimate any correlation between the dependent and independent variables \citep[e.g.][]{Tremaine2002,Kelly2007}. In our case, the independent variables are $a$ = $a(M_\star, P_{\rm orb})$, $M_\star$ and [Fe/H], and they all have measurement uncertainties associated with them. It is necessary to use unbiased estimators rather than OLS. We first experimented with the orthogonal distance regression (ODR) implemented in {\sc scipy.odr} which is described in detail by \citet{odr}. \citet{Kelly2007} derived another Bayesian estimator that accounts for the measurement uncertainty in the independent variables if the independent variables can be described as a mixture of Gaussian distributions. This estimator has been successfully employed in a wide range of astronomical contexts. We coupled the likelihood from the estimators to {\sc MultiNest} \citep{Feroz} for both model selection and sampling parameter posterior distribution. We imposed uniform priors [-10, 10] on the parameters log($\Sigma_0$), $\Delta$log($\Sigma_0$), $l$, $m$ and $n$. We ran {\sc MultiNest} with default settings. Both estimators gave consistent results. We chose to report the results from ODR for its simplicity and because \citet{Kelly2007}'s assumption of Gaussian-mixture independent variables does not strictly apply in this case. The results are presented in Tables 1 to 3 and discussed the next Section.

\begin{figure*}
\begin{center}
\includegraphics[width = 1.\columnwidth]{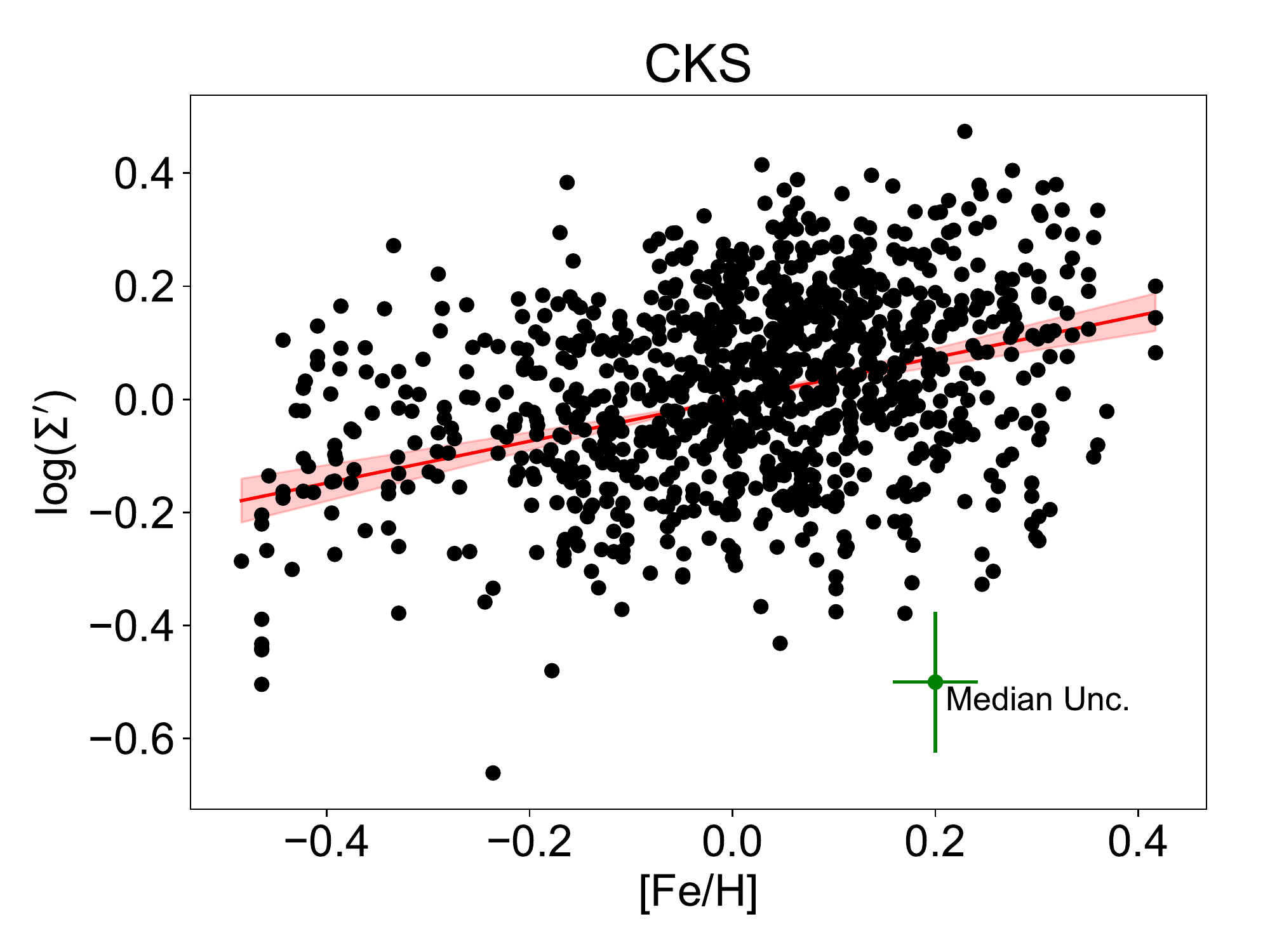}
\includegraphics[width = 1.\columnwidth]{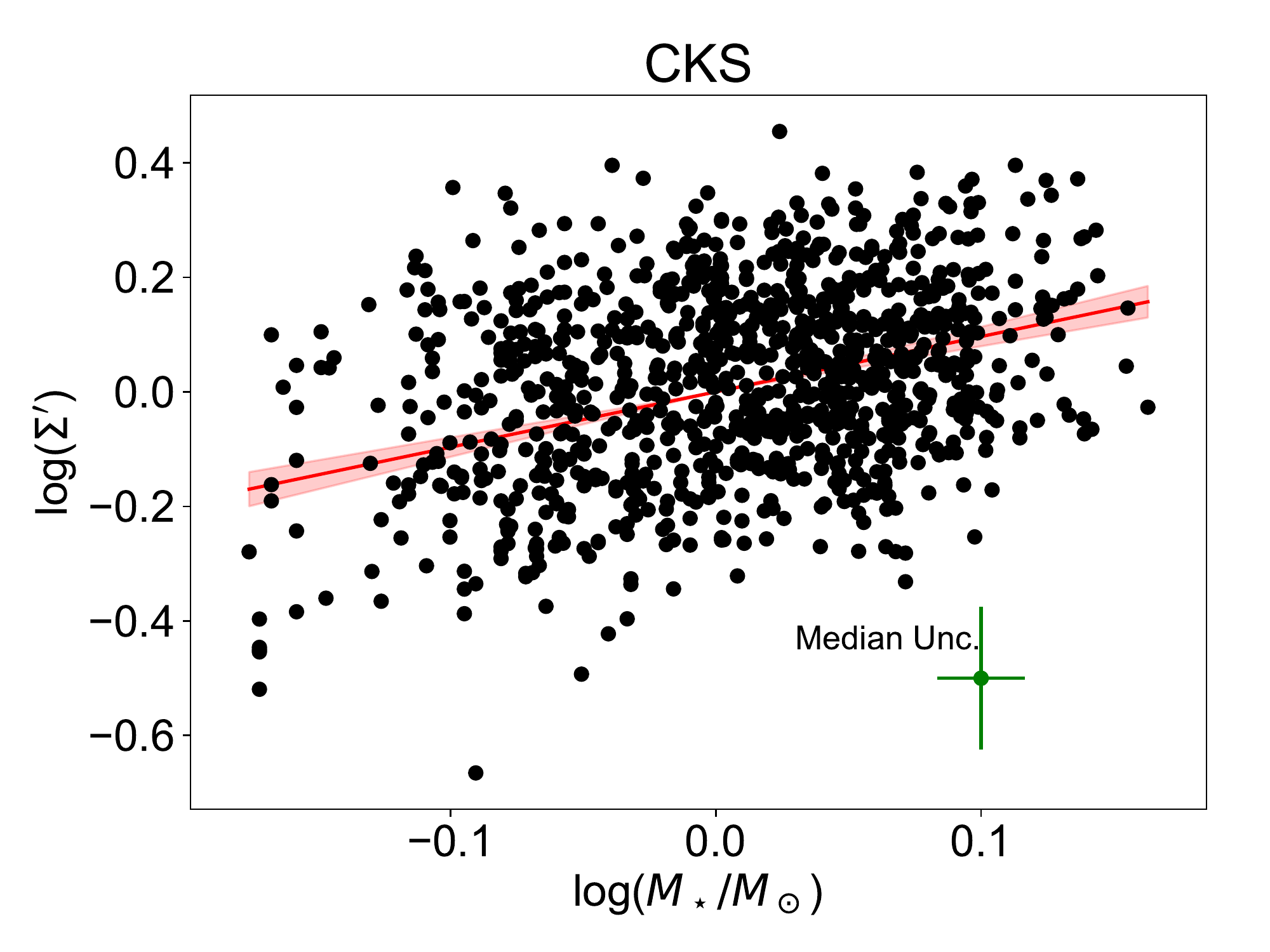}
\includegraphics[width = 1.\columnwidth]{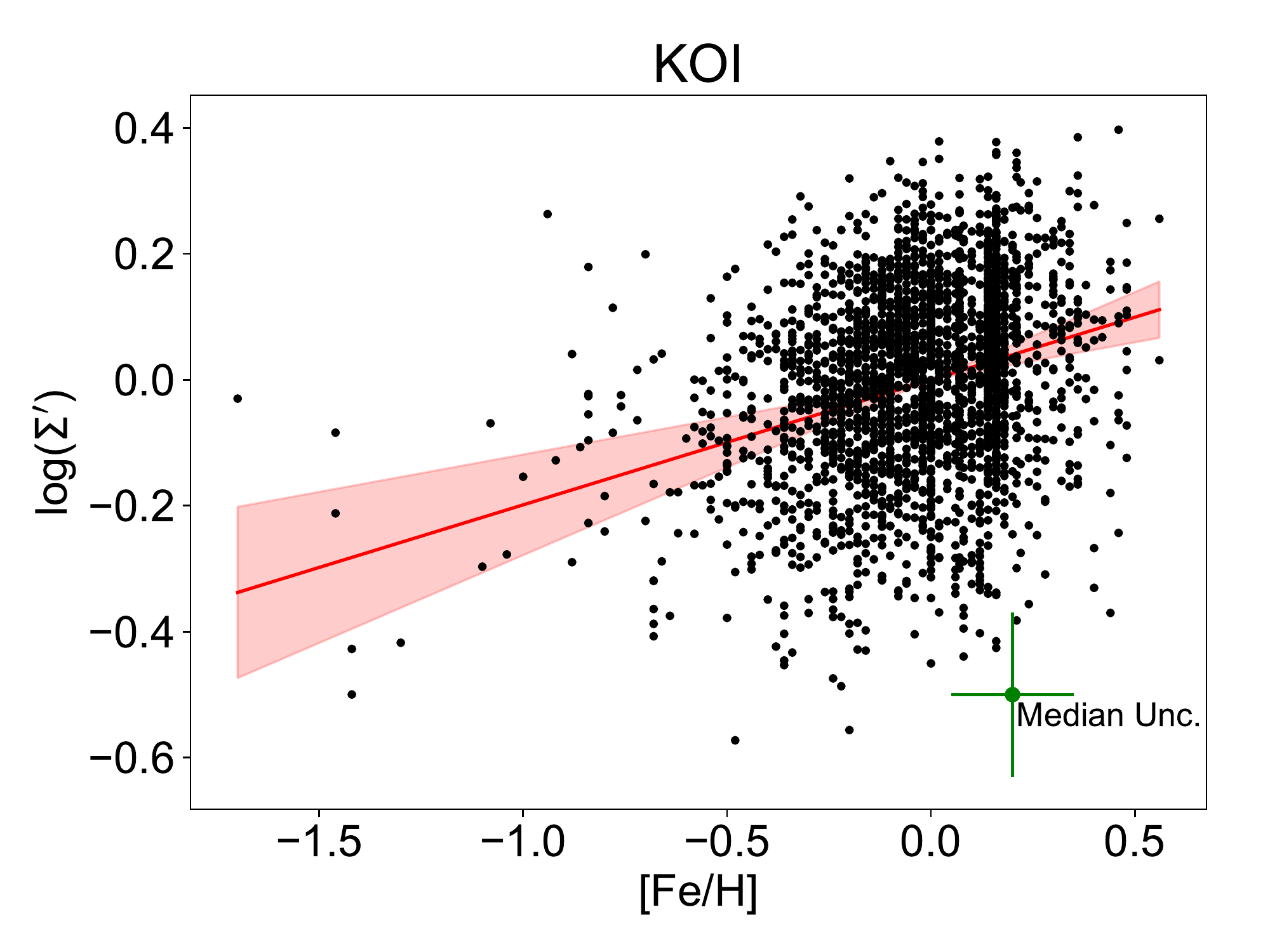}
\includegraphics[width = 1.\columnwidth]{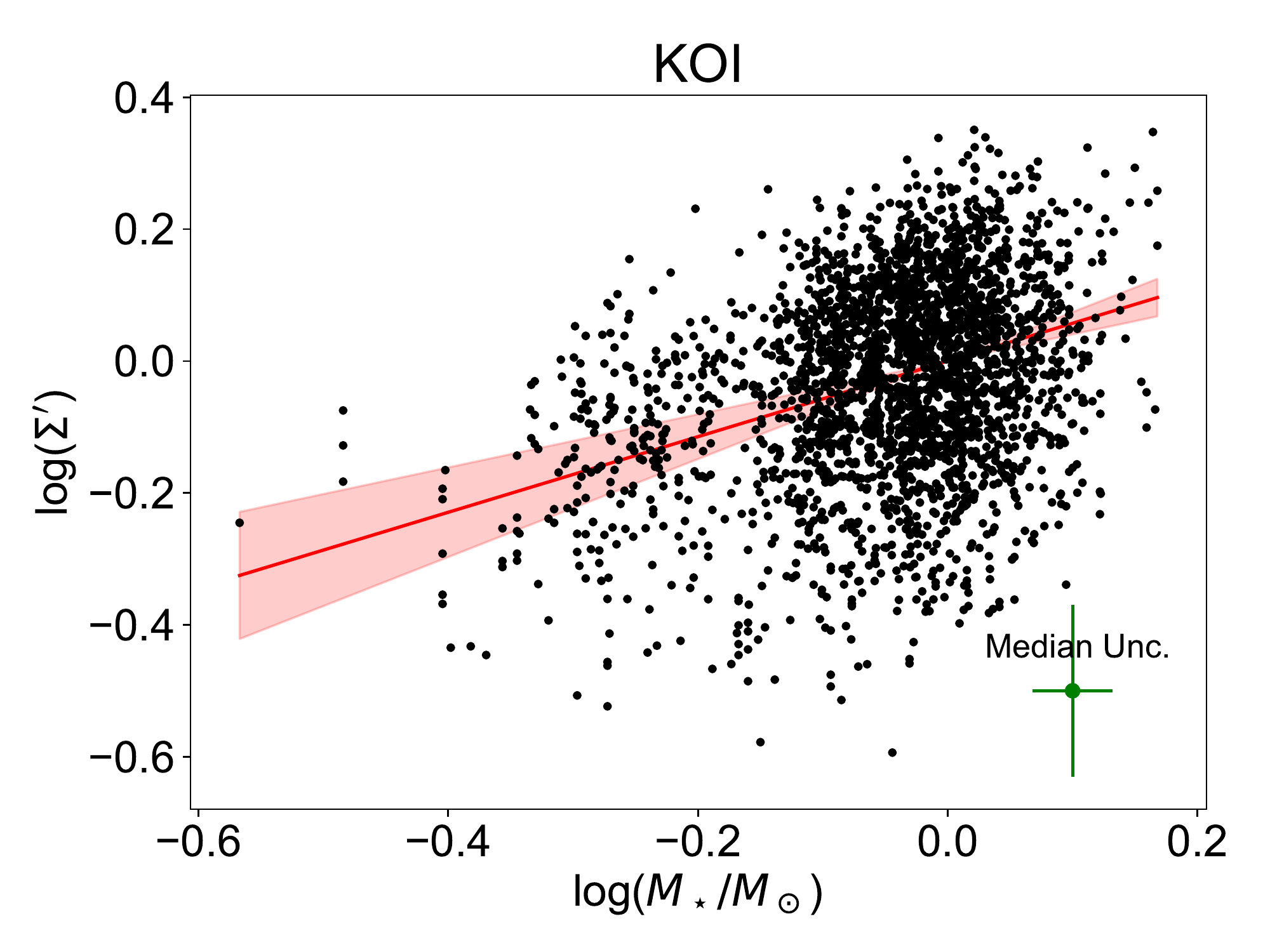}
\caption{The solid surface densities after removing the dependence on orbital distance ($\Sigma^\prime \equiv \frac{\Sigma} {\Sigma_{\rm 0} (a/\rm{AU})^{-1.75}}$) are plotted against the host star mass and metallicity. The errorbars indicate the median uncertainty levels. The red lines show the best-fit power law dependence on [Fe/H] and $M_\star$. The shaped areas represent the 1-$\sigma$ confidence regions. The plots are reproduced for both the CKS and KOI samples.}
\label{fig: correlation}
\end{center}
\end{figure*}

\begin{figure*}
\begin{center}
\includegraphics[width = 1.0\columnwidth]{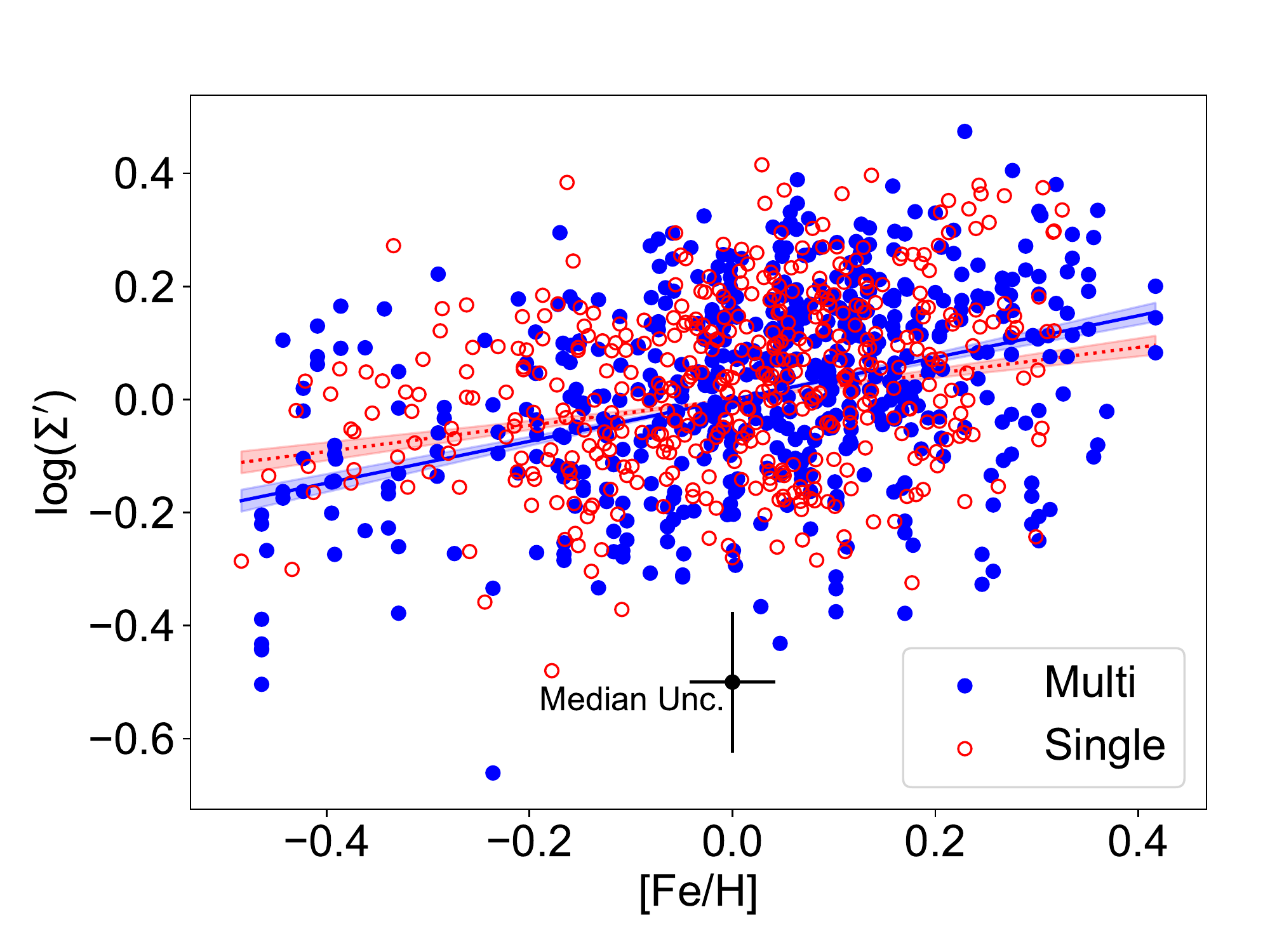}
\includegraphics[width = 1.0\columnwidth]{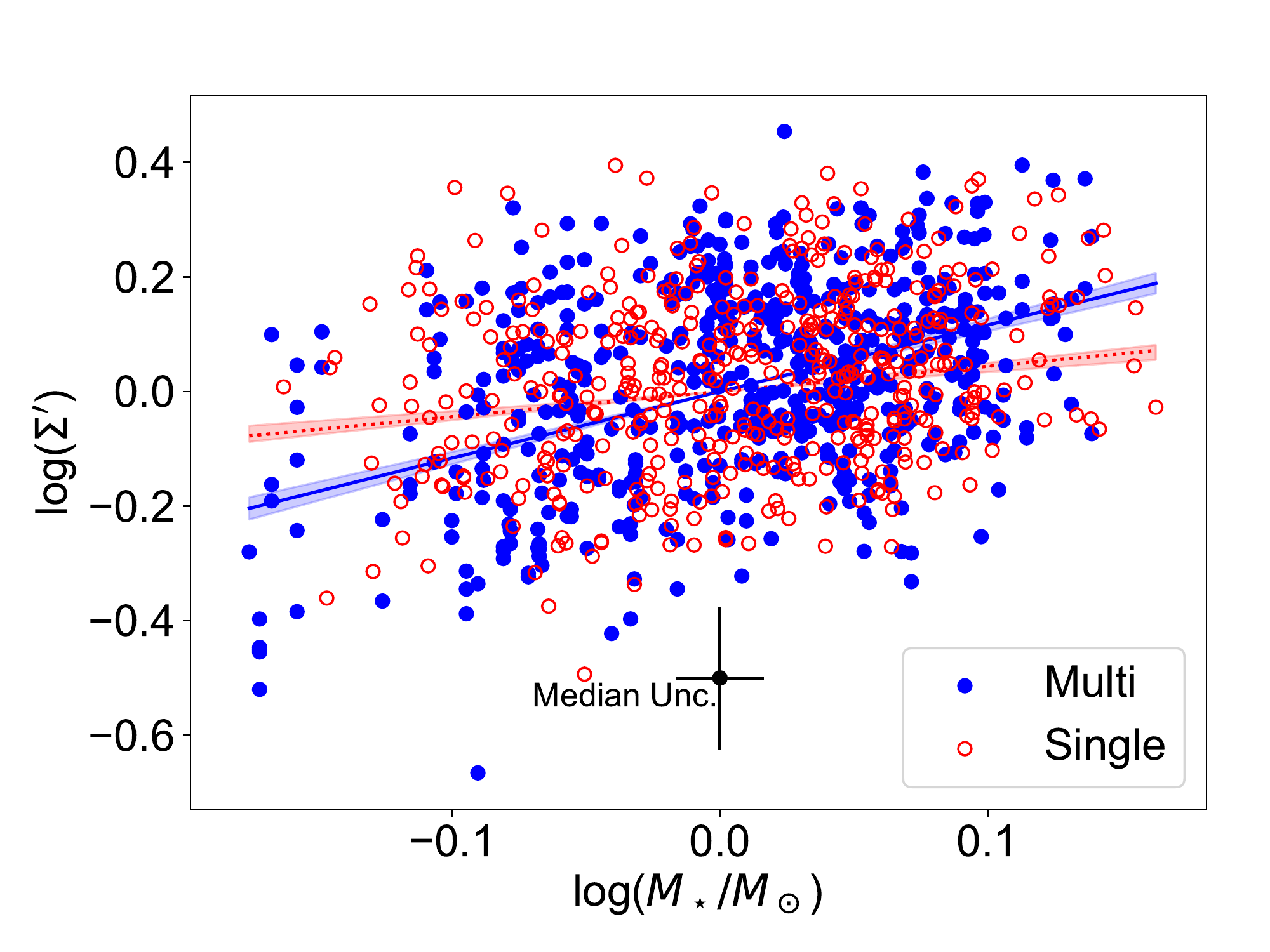}
\caption{Same as Fig. \ref{fig: correlation} after splitting into single-transiting (open red symbols) and multi-transiting systems (filled blue symbols) in CKS. We found that the solid surface density of single-transiting systems showed a much weaker correlation with $M_\star$ and [Fe/H] compared to the multi-transiting systems: 1) the correlation slopes are much weaker (red dotted line versus blue solid lines); 2) the statistical significance according to the marginalized Bayesian evidence $\Delta$log($Z$) from {\sc MultiNest} is several orders of magnitude stronger in multi-transiting systems (>5$\sigma$ v.s. 2$\sigma$; see also Tab. \ref{tab:single_multi}). Our interpretation of this trend is that the architecture of multi-transiting systems suggests a dynamically quiet history which largely preserved the imprints of formation out of the natal disk. On the other hand, the dynamically hot history (giant impact collisions, planet-planet scattering etc) that gave rise to single-transiting systems partially removed these imprints.}
\label{single_vs_multi}
\end{center}
\end{figure*}

\section{Results and Discussion}
\subsection{Strong $M_\star$ but weak [Fe/H] Dependence}
\label{sec:strongMs}

Previous works made contradictory claims on whether $M_\star$ or [Fe/H] leave a stronger imprint on the observed distribution of sub-Neptune planets \citep[c.f.][]{OwenMurray,Wu}. Using the location of the radius gap as a function of host star properties, \citet{Wu} argued for a linear scaling between planetary mass and stellar mass and no correlation between planet size and host star metallicity. However, \citet{OwenMurray} seem to disagree. They found that stellar metallicity is correlated planetary radius (mass). More importantly, they showed for planets with $<$2.5-day orbits (where one expect complete photoevaporation) and planets with $>$25-day orbits (where one expect limited photoevaporation), it is still the case that higher-metallicity stars host larger planets. They suggest that some effects, other than the metallicity-dependent efficiency of photoevaporation, must be at work in deciding the size of sub-Neptune planets. 

Here we offer an alternative, also more quantitative perspective using the MMEN framework. One caveat before going to the results is that the stellar mass $M_\star$ and metallicity [Fe/H] in the CKS sample are strongly correlated with each other (Fig. \ref{ms_vs_met}). Previous works have attributed this covariance to Galactic chemical evolution i.e. more massive stars are younger and hence they likely formed from the more metal-enriched Galaxy \citep{Fulton2018}. However, several works directly looked for a possible correlation between age and metallicity in the young thin disk stars; and did not find a compelling trend \citep{Bensby,SilvaAguirre2018}. While investigating this covariance further, we found that it is also related to sample selection and model degeneracy. In Fig. \ref{correlation}, we plotted the effective temperature $T_{\rm eff}$ and mass $M_\star$ of the CKS and KOI stars while colorcoding the points with metallicity [Fe/H]. At the same stellar mass, a lower metallicity translates to lower opacity in the stellar atmosphere. The star would appear smaller but hotter. The variation of [Fe/H] by 1 dex results in a change of $T_{\rm eff}$ by hundreds of kelvins. The CKS sample was constructed with a hard boundary on the effective temperature 4700K$<T_{\rm eff}<$6500K, therefore the low-metallicity stars at the higher-mass end (>1.2$M_\star$; [Fe/H]<0) were excluded by the survey. An analogous effect occurs at the lower-mass end. These effects contribute to the $M_\star$-[Fe/H] covariance in the CKS sample.

We tried to disentangle the influence of $M_\star$ and [Fe/H] by modeling them separately in Eqn \ref{eqn:powerlaw} and used the Bayesian evidences log(Z) computed by {\sc MultiNest} to gauge which of them has a stronger effect. The parameter posterior distribution and the Bayesian evidence as inferred from {\sc MultiNest} are summarized in Tables 1 to 3. We can see that $M_\star$ generally has much stronger explanatory power in the variation of MMEN $\Sigma$ compared to [Fe/H]. This is manifested by the steeper correlation slope $m<n$. Furthermore, adding the $M_\star$ dependence led to several orders of magnitude improvement in the Bayesian evidence Z compared to the [Fe/H] dependence. For a summary of the results, we calculated the average MMEN profile
with the CKS multi-transiting planets thanks to its uniformity and dynamical quietness (we will discuss why we excluded single-transiting systems in Section \ref{sec:single_multi}): 

\begin{equation}
\label{powerlaw_result}
\begin{split}
\Sigma= 50^{+33}_{-20} \rm{~g~cm}^{-2} (\frac{a}{\rm AU})^{-1.75\pm 0.07}\\
~10^{0.22\pm0.05 {\rm [Fe/H]}}~(\frac{M_\star}{M_\odot})^{1.04\pm0.22}
\end{split}
\end{equation}

\noindent We note that the systematic variation between different mass-radius relationships outweighs the internal uncertainties from sampling the parameter posterior distribution (See Table 1 to 3). To capture this systematic uncertainty, we report here the MMEN profile using the mean and the standard deviation of the model parameters inferred from different mass-radius relationships (Section 3.1). 

This strong dependence on $M_\star$ and the weaker dependence on [Fe/H] are seen in the KOI and RV samples too (Table 1 and 3). Moreover, we restrict our attention to Sun-like stars (0.9$M_\odot<M_\star<1.2M_\odot$; -0.2<[Fe/H]<0.2) where there is limited covariance between $M_\star$ and [Fe/H] (Fig. \ref{correlation}); and repeated the analysis. We arrived at a result similar to Eqn \ref{powerlaw_result}, although the smaller dynamical range weakens the statistical significance. As another check of the applicability of Eqn \ref{powerlaw_result} across different stellar types, we repeated the MMEN analysis for the TRAPPIST-1 system: one of the most characterized, low stellar mass planetary system \citep[0.09$M_\odot$, ][]{trappist}. We arrived at a solid disk surface density $\Sigma\approx 6{\rm{~g~cm}^{-2}}$ (see Fig. \ref{fig:all_systems}) at 1AU that is about 10\% of that Eqn. \ref{powerlaw_result}, confirming the linear dependence on $M_\star$.

These observations gave us more confidence that $M_\star$ plays a more important role in the MMEN framework. The $\Sigma\propto M_\star^{1.04\pm0.22}$ dependence fits the expectation of a simple {\it in-situ} formation scenario. Millimeter observations of the dust continuum by \citet{Andrews} revealed that the dust mass of a protoplanetary disk scales more or less linearly with host star mass: $M_{\rm dust} \propto M_\star$. It is perhaps not surprising that this linear relation continues to the <1AU innermost disk where {\it Kepler}-like planets reside: $\Sigma\propto$ $M_{\rm dust} \propto M_\star$.

Invoking the more sophisticated $\alpha-$disk model and assuming steady accretion in the inner disks, the total surface density ($\Sigma_{\rm tot}$) is linked to mass accretion rates ($\dot{M}_{\rm acc}$) by   
\begin{equation}
\dot{M}_{\rm acc}\propto\alpha c_{\rm s}^2 \Sigma_{\rm tot}\Omega^{-1}
\end{equation}
\noindent where $\alpha$ is the viscous parameter, $c_{s}$ is the sound speed and $\Omega$ is the Keplerian angular velocity. The sound speed $c_{s}$ is  related to the gas temperature ($T_{\rm gas}^{0.5}$) in the inner disk regions with 
\begin{equation}
T_{\rm gas}^4=T_{\rm acc}^4+T_{\rm irr}^4
\end{equation}
\noindent where $T_{\rm acc}$ is  due to heating from the viscous dissipation, and $T_{\rm irr}$ is due to the stellar irradiation. If the viscous dissipation is the dominant source
of heating, $T_{\rm gas}^4\propto T_{\rm acc}^4\propto M_{\star}\dot{M}_{acc}$ at a given disk radius \citep{Hartmann2018}, and $\Omega\propto M_{\star}^{0.5}$. Furthermore,  from the observations made by \citet{2009ApJ...703..922V,2017A&A...600A..20A}, $\dot{M}_{\rm acc}\propto M_{\star}^{1.3\pm0.3}$ for  $0.2M_{\odot}\leq M_{\star}<3M_\odot$. Thus, $\Sigma_{\rm tot}\propto M_{\star}^{1.2\pm0.3}$. Conversely if stellar irradiation dominates disk heating, $T_{\rm irr}^4\propto R_{\star}L_{\star}$ at a given disk radius \citep{2007prpl.conf..555D}, where  $R_{\star}$ and $L_{\star}$ are stellar radius and luminosity respectively of the central young stars. For 1-3 Myr-old stars, pre-main evolutionary models give $L_{\star}\propto M_{\star}^{1.5}$ and $R_{\star}\propto M_{\star}^{0.4}$ for 0.6\,$M_{\odot}$$\leq M_{\star}\leq$1.4\,$M_{\odot}$ \citep{2000A&A...358..593S}, which leads $\Sigma_{\rm tot}\propto M_{\star}^{1.3\pm0.3}$. Now if we marginalize over [Fe/H], the MMEN solid surface density inferred from CKS multi-planet systems has a profile quite close to the predictions of $\alpha$-disk models:

\begin{equation}
\begin{split}
\Sigma= 50^{+32}_{-19} \rm{~g~cm}^{-2} (\frac{a}{\rm AU})^{-1.76\pm 0.07}~(\frac{M_\star}{M_\odot})^{1.39\pm0.29}
\end{split}
\end{equation}

If the {\it in-situ} planet formation is only limited by the availability of solid materials in the disk, one might also expect $\Sigma\propto 10^{{\rm [Fe/H]}}$ i.e. assuming disk solid mass scales with the host star metallicity. Alternatively, if {\it in-situ} planet formation is limited by the coagulation rate of planetesimals, one might expect a steeper dependence e.g. $ \propto 10^{2{\rm [Fe/H]}}$ often seen in collisional problems. This latter scenario is often invoked to explain the strong correlation between giant planet occurrence and host star metallicity \citep{Fischer2005}. However, we only found a sub-linear relation $\Sigma\propto 10^{0.22\pm0.05{\rm [Fe/H]}}$. This relation confirms the picture that the formation of sub-Neptune planets occurs readily in lower metallicity ([Fe/H] $\approx$ -0.4) systems while the formation of giant planets strongly favors metal-rich environments \citep[]{CKS4}.  As we mentioned in Section 2, our analysis excluded gas giants because we could not estimate the solid mass accurately with only the planetary radius. By excluding the giant planets (>4$R_\oplus$), we have excluded the most massive cores from this analysis and thus weakened the correlation between MMEN $\Sigma$ and [Fe/H]. On a related note, \citet{Zhu_metallicity} found that the fraction of stars with sub-Neptune planets increased steadily from low to high [Fe/H] systems, whereas the average number of sub-Neptune planets per star increases with [Fe/H] initially but plateau at [Fe/H]$\sim0.1$ possibly caused by the emergence of giant planets in higher-metallicity systems. Could the emergence of giant planet dynamically disrupt its sub-Neptune companions? We will discuss this point in more detail in the next Section.

\subsection{Singles v.s. Multis}
\label{sec:single_multi}
Several observations indicate that the orbital architectures of {\it Kepler} single-transiting and multi-transiting systems are distinct. \citet{Xie}, \citet{vaneylen} and \citet{Mills2019} arrived at consistent conclusions that single-transiting systems tend to have more eccentric orbits while multi-transiting planets favor circular orbits. \citet{Fang2012} and \citet{Zhu2018} showed that systems with fewer planets have substantially larger mutual inclination dispersion.  \citet{Moriarty} attributed this architectural difference to different disk properties of the single versus multi-planet host stars. However, later work by \citet{Weiss2018b} showed that singles and multis are very similar in terms of both planetary and stellar properties betraying a common origin. Another possible explanation for the single-multi difference is the dynamical interaction between the sub-Neptune planets themselves or with giant planets in the same system. \citet{ZhuSE-CJ} and \citet{Bryan} showed that the occurrence of {\it Kepler}-like sub-Neptune planets and cold Jupiters (>1AU) are correlated: as much as 30\% of {\it Kepler}-like systems has a cold Jupiter whereas a cold Jupiter almost certainly has inner {\it Kepler}-like planetary companions. A follow-up study by \citet{Masuda} revealed that the mutual inclination between the cold Jupiter and inner planetary system is drastically larger if the inner planetary system only has one transiting planet. The proposed explanation is that an inclined cold Jupiter dynamically disrupt or scatter the inner planets thereby reducing the number of planets and/or inducing large mutual inclinations and eccentricities. This dynamically hot history significantly modifies the initial orbital architecture of a system which is manifested today as the lower multiplicity, higher mutual inclination and higher orbital eccentricities of the single-transiting systems.

Knowing this potential architectural difference, we split the CKS sample into single-transiting and multi-transiting systems and fitted their MMEN separately. The results are summarized in Tab. 2. The slopes of correlation between $\Sigma$, [Fe/H] and $M_\star$ ($m$ and $n$) are weaker in the single-transiting systems (see Fig. \ref{single_vs_multi}). In addition, looking at the Bayesian evidence Z as a statistical measure of model improvement, including [Fe/H] and $M_\star$ dependence in MMEN for single-transiting systems resulted in very limited model improvement (Bayesian evidence Z improved by no more than 2 orders of magnitude, or a 2-$\sigma$ significance). On the other hand, it is more than 15 orders of magnitude for the multi-transiting systems (or >5-$\sigma$ significance). Our interpretation is that the dynamically hot history of single-transiting systems has at least partially eliminated the memory of formation out of the disk. Imagine that both the single-transiting and multi-transiting systems form a gaseous disk. After the disk dissipates, dynamical interaction between the planets or with a distant cold Jupiter start to incite a dynamical upheaval in some of the systems. The dynamically hotter systems have reduced number of planets on more eccentric, mutually inclined, and scrambled orbits. These systems are naturally observed as single-transiting and they have also lost the memory of the initial orbital architecture from {\it in-situ} formation. Therefore the singles poorly fit a MMEN framework. On the other hand, the dynamically colder systems are more likely observed as multis and largely preserved the initial orbital architecture.  With the MMEN framework, we have rediscovered the different dynamical history that underpins the observed architectural differences between the single-transiting and multi-transiting systems \citep{Xie,vaneylen,Mills,Zhu2018,Weiss2018b}.
%

\subsection{Occurrence Rate Decline $<$0.1AU: Disk Thinning or Truncation}\label{sec:innerdisk}
A prominent feature displayed by sub-Neptune planets is that the occurrence rate as a function of orbital distance follows a broken power law: the occurrence rate increases steadily with orbital distance until about 0.1AU (orbital period of $\sim10$ days) after which the occurrence rate plateaus out to 1AU across FGKM hosts \citep[e.g.][]{Dressing,CKS4}. In Fig. \ref{occurrence}, we reproduced this result using the CKS sample. 

 The coagulation rate of planetesimals cannot be blamed for this decline of occurrence rate because the coagulation rate increases with orbital velocities hence higher in the inner disk \citep{CL2013}. One possible explanation of the occurrence rate decline $<$0.1AU is that the amount of planet-forming solid materials may be significantly reduced at the inner disk due to dust sublimation, rapid radial drift of dusts or some other processes. We call this possible scenario ``Solid Disk Thinning''. If this were true, the inner disk should host fewer and smaller planets; and one should observe a corresponding drop in the MMEN solid surface density $\Sigma$ within 0.1AU. We note that occurrence rate declined by about 1.5 orders of magnitude, a corresponding change in MMEN $\Sigma$ would have been visually obvious in Fig.\ref{a_vs_sigma}, but was not seen. 
 
\begin{figure}
\begin{center}
\includegraphics[width = 1.0\columnwidth]{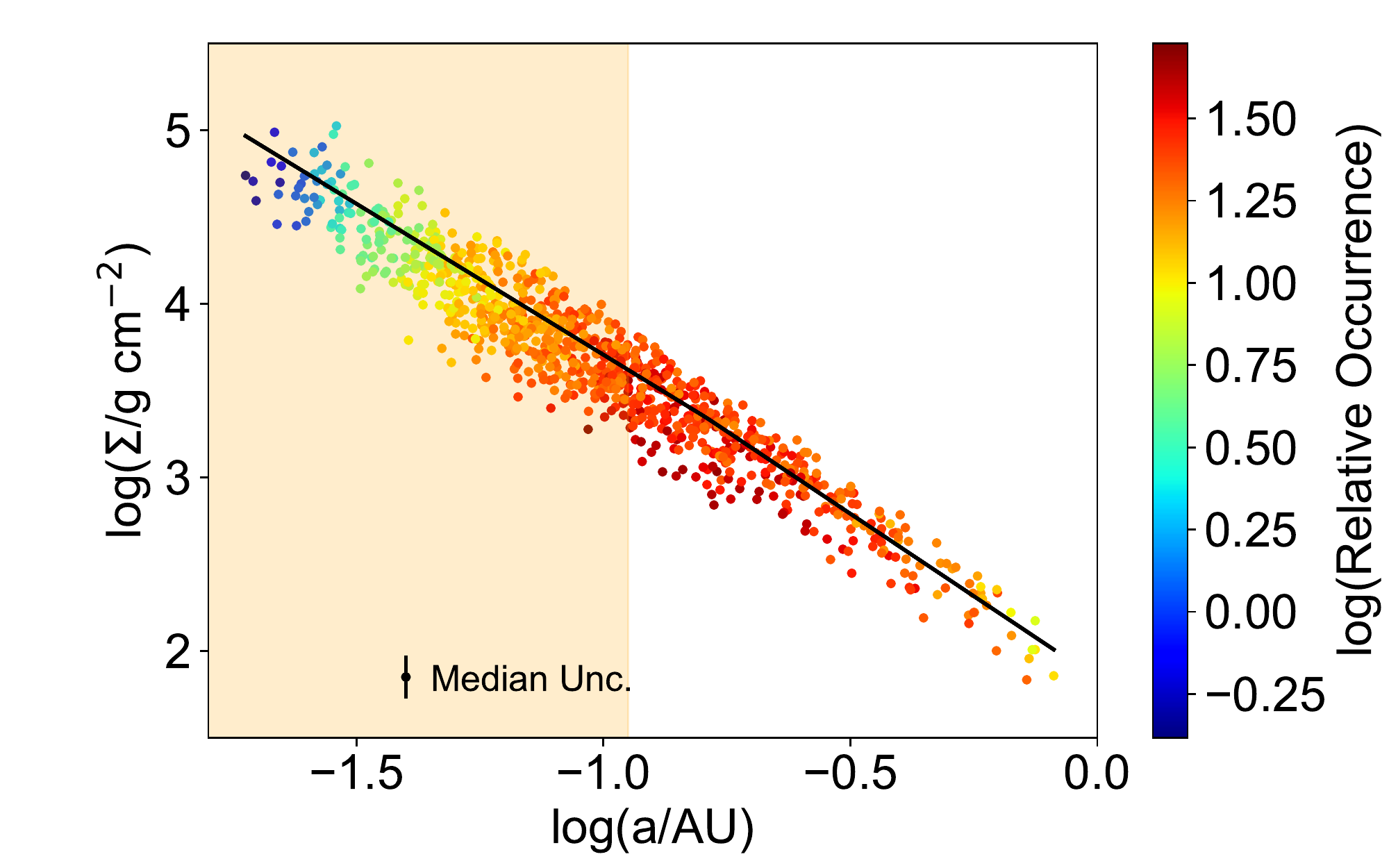}
\caption{Same as Fig. \ref{a_vs_sigma}. The orange shaded region represent the innermost region of the protoplanetary disk where a steep decline of sub-Neptune occurrence (color-coding) by about 1.5 order of magnitude has been observed \citep[see Fig. \ref{occurrence} and previous works e.g. ][]{CKS4}. If the occurrence decline is due to the sublimation of the solid disk at the innermost region, one would expect a significant decline in the MMEN solid surface density $\Sigma$. We fitted a broken power law model for the solid surface density $\Sigma$ as a function of orbital distance $a$ (black lines are random posterior samples). The broken power law is not statistically favored by the data; rather $\Sigma$ follows a smooth profile throughout the innermost 1AU of the disk. This favors the "Truncation" rather than "Sublimation" of the inner disk as the driver for the decline of planet occurrence. See Section \ref{sec:innerdisk} for detail. }
\label{fig:broken_power_law}
\end{center}
\end{figure}

 We also performed a more quantitative test. We compared two models: 1) a single power law of $\Sigma$ as a function of orbital distance $\Sigma\propto a^{l}$; and 2) a broken power law where the power law index $l$ is allowed to switch to a different value below a critical distance $a_{\rm crit}$. Using the same procedure described in Section 3, we performed model selection using the Bayesian evidence $\Delta$log(Z) computed by {\sc MultiNest}. The broken power law, with the introduction of two additional parameters, was not favored by the data (Fig. \ref{fig:broken_power_law}). This statement is true for any of the mass-radius relationships in Section 3.1. The $a_{\rm crit}$ parameter is also poorly constrained rather than being pinned down to $\approx$0.1AU as in the occurrence rate model. 
 
 We therefore disfavor the ``Solid Disk Thinning'' scenario and argue for the alternative explanation ``Solid Disk Truncation'' \citep{Lee_usp}. In this case, the protoplanetary disk maintains a smooth solid density profile until it is truncated by e.g. the magnetosphere of the host star. In a truncated disk, planets can form {\it in-situ} relatively undisturbed beyond the truncation radius. On the other hand, planet formation is totally quenched within the truncation radius. Since truncation occurs the stellar co-rotation radius, the rotation period of the host star sets the inner edge of planet formation. For an ensemble of stars, one therefore observes a decline of occurrence rate towards the host star. On the other hand, the MMEN disk profile, thanks to closer-in planets around faster-rotating stars, remains smooth. 
 
 In short, the "Solid Disk Thinning" scenarios should manifests as a significant drop of $\Sigma$ within 0.1AU i.e. the upper left corner of Fig. \ref{fig:broken_power_law}; while "Solid Disk Truncation" should give rise to fewer planets in the upper-left corner that lie around a continuous trend from outer disk. This truncation scenario was discussed in detail by \citet{Lee_usp}. Their magnetosphere truncation scenario can very well reproduce the observed broken power law occurrence rate profile with the observed rotation periods of young stars in clusters and the equilibrium tidal decay of planets after disk dissipation. We note that the disk truncation creates a pressure bump in the inner disk and the resultant inverted pressure gradient prevents the rapid loss of dusty materials due to radial drift. This maintains the solid density in the inner disk and also allows more time for the formation of close-in sub-Neptune planets.

\subsection{RV and TTV Samples}
\label{rv-ttv}
The main conclusions of this paper are drawn from analyzing the uniform and precise CKS sample. However, since most CKS planets do not have measured masses, we had to rely on mass-radius relationships. We repeated our analyses on systems with direct planetary mass measurements: 145 TTV planets in 55 systems from \citet{Hadden} and 135 RV planets in 51 systems from {\sc Exoplanet Archive} (see Fig. \ref{fig:all_systems}).

One word of caution before going into the results of the RV sample is that we do not have a simple way to quantify detection bias. The RV planets were derived from a host of RV surveys with different target selection criteria, observation strategies and instrument characteristics. Rather than embarking on a tour de force of quantifying the detection bias ourselves, we simply left the detection bias of the RV planets uncorrected in this work. Nonetheless, the inferred MMEN profile from RV planets is very similar to that of the CKS multi-planets: 
\begin{equation}
\label{RV_results}
\begin{split}
\Sigma= 100^{+90}_{-48} \rm{~g~cm}^{-2} (\frac{a}{\rm AU})^{-1.77\pm 0.11}\\
~10^{0.07\pm0.16 {\rm [Fe/H]}}~(\frac{M_\star}{M_\odot})^{0.87\pm0.23}
\end{split}
\end{equation}

\noindent i.e. a stronger almost linear correlation with $M_\star$ but a weaker dependence on [Fe/H]. We also note that the inferred solid surface density is somewhat higher in the RV sample compared to the CKS sample. The mean solid disk density is about 2$\sigma$ higher in logarithmic space: log($\Sigma_0/{\rm g~cm}^{-3}$) = $10^{1.70\pm0.06}$ v.s. $10^{1.99\pm0.10}$ (Tab. 3). One may be tempted to attribute this difference to a detection or publication bias towards heavier planets in RV surveys \citep{Mills}. We join previous authors \citep{Burt,Montet} in urging the community to publish upper limits mass constraints for statistical unbiased analyses.

For the TTV sample, we did not robustly detect the $M_\star$ or [Fe/H] dependence of MMEN. The model favored by Bayesian evidence comparison only contains a dependence on orbital distance: $\Sigma= 72^{+30}_{-21} \rm{~g~cm}^{-2} (\frac{a}{\rm AU})^{-1.69\pm 0.09}$. The simple explanation is that the much smaller sample size (145 planets) and limited dynamical range in the TTV sample was not able to reveal the more subtle $M_\star$ and [Fe/H] dependence. However, this is probably not the whole story. The RV sample contains a similar number of planets yet the Eqn. \ref{RV_results} agreed well with the CKS sample Eqn. \ref{powerlaw_result}. We performed a bootstrap test by randomly selecting a sample of 145 planets from the CKS sample and repeating the MMEN analysis. We found that the size of the 145 TTV planets should have been more than sufficient to recover the $M_\star$ or [Fe/H] dependence of MMEN as in Eqn. \ref{powerlaw_result}. Our boostrap analysis gives a Bayesian Evidence improvement of $\sim$7 orders of magnitudes or 4-$\sigma$ confidence. We thus provide an interesting but more speculative explanation that involves a different formation pathway of TTV planets. TTV signals are preferentially measured for planets near mean-motion resonances. It is proposed that these near resonant systems formed further out in the protoplanetary disk followed by convergent disk migration that locked the planets into resonance\citep{Lee,MillsNature}. The colder outer disk facilitates the accretion of gaseous envelopes and partially explains the observed inflated radii of planets near resonance \citep{Millholland2019}. With such a formation pathway, TTV planets must have undergone some amount of disk migration; they are misplaced in the MMEN frame which is an inherently {\it in-situ} formation idea. The TTV planets simply do not encode the disk properties at their current-day orbits; it is hence not surprising that the $\Sigma$ inferred from TTV planets do not show $M_\star$ and [Fe/H] dependence.

\subsection{Higher Formation Efficiency for Lower-mass Stars}
Another interesting observation is that the occurrence rate of close-in sub-Neptune planets increases steadily towards lower mass stars \citep{Dressing,Mulders2015}. In Fig. \ref{occurrence} we reproduced this result using the CKS sample. More intriguingly, \citet{Mulders2015b} showed that the total solid mass locked in planets also increases steadily towards lower mass host stars. For FGKM stars, the amount of solid mass in planets are respectively $3.6 \pm 0.1 M_\oplus$, $5.0 \pm 0.1M_\oplus$, $5.4 \pm 0.2M_\oplus$ and $7.3 \pm 0.7M_\oplus$. Here \citet{Mulders2015b} directly summed the mass of detected {\it Kepler} planet multiplied by a correction factor equal to the inverse of the detection probability and transit probability. At this point, the readers may be confused: how does this result compares to our MMEN analysis? Briefly, our MMEN analysis spreads the mass of each planet into its feeding zone to probe the local disk surface density. We then fitted the profile of surface density with a power law assumption. \citet{Mulders2015b} strictly only counted the amount of solid mass that is locked in existing planets, whereas the MMEN framework uses planets as anchor points to sample the solid disk profile in a {\it in-situ} formation setting. 

With that in mind, we can now calculate the efficiency of planet formation. We integrated the solid disk profile (Eqn \ref{powerlaw_result}) from $0.02$AU to 0.7AU  \citep[the boundaries are chosen to match those of][]{Mulders2015b}. We found that for FGKM stars the average total solid mass in the inner disk are $27M_\oplus$, $23M_\oplus$, $19M_\oplus$ and $11M_\oplus$ subject to a scatter of about 0.2 dex. These translate to a {\it in-situ} formation efficiency of $3.6/27 \approx 13\%$, $5.0/23 \approx 22\%$, $5.4/19 \approx 28\%$, $7.3/11 \approx 67\%$ for FGKM stars. Alternatively, one can also think of this ratio as a fraction of surviving planets after dynamical evolution. In both cases, this fraction increases monotonically towards lower mass stars. We caution that the result for the M stars is mostly based on by the RV sample (Eqn \ref{RV_results}); the more uniform and more precise CKS sample has no M stars due to difficulties in modeling M star atmospheres. This situation will be improved with the CKS-Cool project (Petigura et al. in prep) that extends to lower-mass stars using more empirical spectral methods.  One caveat of this result is that the width of feeding zone may somehow depend on the spectral type of the host star. If true, this would bias the inferred MMEN surface density. Moreover, we have also assumed the same innermost radius of the disk and the same radial dependence ($l$) for different stellar types. The first assumption on innermost radius can be partially justified by the fact that planets on extremely short orbital periods ($\approx 0.02$AU) have been observed across FGKM stars \citep{USP}. We repeated our MMEN analysis in Section 3 for stars of different spectral types separately. We found that radial dependence ($l$) is similar across stellar types as we assumed; it shows a standard deviation of 0.06 consistent with the estimated uncertainty in Eqn. \ref{powerlaw_result}.

\citet{Yang} analyzed in conjunction the occurrence and architecture (multiplicity and mutual inclination) of {\it Kepler} planets following the methods of \citet{Zhu2016}. They found that for late-K, early-M stars (<4000K) the fraction of stars hosting sub-Neptune planets is $\sim$55\% with a typical intrinsic multiplicity of 3 planets, whereas the numbers drops steadily to early-type stars ($\sim$15\% F stars host {\it Kepler}-like systems with a multiplicity of $\sim$2). This is additional evidence for the higher efficiency of forming/retaining planets for lower-mass stars. What could be the underlying reason? We know that M dwarfs are much less likely to host a giant planet \citep{Cumming,Clanton}. For giant planets within $<2000$ days period, the suppression factor is about 3-10 times fewer than sun-like stars. \citet{Masuda} showed that the presence of a misaligned giant planet dynamically disturb the inner planetary systems possibly reducing the number of planet and inducing higher mutual inclinations. The lack of giant planets around lower-mass stars may be a boon for the close-in sub-Neptune planets which are much likely to survive around M dwarfs.

A recent work by \citet{Moe} showed that stellar companions with $a<1$AU completely suppresses the formation of S-type planets around the primary, while those at about 10AU suppresses planet occurrence to 15\% that of single stars. \citet{Moe} further argues that the higher occurrence rate of sub-Neptune planets around lower-mass stars can be partially accounted for by the corresponding lower binary fraction of their hosts.

Another explanation for higher efficiency of planet formation around lower-mass stars is their longer disk lifetimes. By examining the existing disk fraction across clusters of different ages, \citet{Kennedy} suggested that lower-mass stars have longer lasting disks. \citet{Silverberg} reported several ''Peter Pan'' disks around $> 20$ Myr M dwarfs. \citet{Lee2019} showed that the disk lifetime is a major factor in the assembly of planetesimals into planetary cores. The longer disk lifetime of lower-mass stars might be instrumental for converting a larger fraction of the disk solids into sub-Neptune planets.

In a separate line of argument, there is a whole literature on the possibility of planet engulfment after planet formation \citep[see the review by][]{Ramirez}. Comoving binaries are often assumed to be coeval and chemically similar, however detailed abundance studies revealed that sometimes one of the star can be preferentially enhanced in refractory materials e.g. $\sim$ 0.2 dex \citet{Oh}. A possible explanation is the ingestion of $\sim15M_\oplus$ worth of Earth-like materials in the star's convective layer. Planet formation and evolution can be an intrinsically lossy process that a significant fraction of disk solid materials do not end up in planets. We note that so far the detected possible engulfment signatures are all around G and F-type stars \citep{Oh,Ramirez,Nagar}. However, the existing engulfment surveys are far from being uniform across stellar types, it is hence premature to claim whether planet engulfment happens more frequently around early-type stars.

\subsection{Is There a Universal MMEN?}
\label{sec:universal}

\citet{Raymond} constructed the MMEN using {\it Kepler} multi-transiting systems with 3-6 transiting planets. They used the geometric means of planets' semi-major axes as the boundaries of the feeding zones (see Section 3.2). This prescription is very susceptible to non-transiting planets. Take the Solar System for example, it is very unlikely, even with infinite observation time, that any external observer could see three planets transit his/her line of sight \citep{Wells}. The MMEN this observer derives with the prescription of \citet{Raymond} would be drastically different from the MMSN one derives knowing the complete planetary inventory of the Solar System. Applying their prescription, \citet{Raymond} found that the MMEN profile for individual {\it Kepler} systems can range between $\Sigma$ $\propto a^{-3.2}$ to $a^{0.5}$. To correct for the effect of missing planets, \citet{Raymond} used a 6-planet model with 5-degree dispersion in mutual inclination to mimic {\it Kepler} observations. Their model failed to obtain a satisfactory fit to the observed {\it Kepler} systems. They hence cast doubt on a universal MMEN and argued against {\it in-situ} formation. As we mentioned in Section 3.2, using the geometric mean of planets' semi-major axes as feeding zone is particularly susceptible to missing planets. A missing planet produces a large gap in orbital separation which in turn produces anomalously low surface densities for its neighbours. Moreover, a 5-degree mutual inclination is simply too restrictive, recent works showed that the mutual inclination distribution depends on both multiplicity and orbital distance \citep{Zhu2018,Dai}. 

\citet{Schlichting2014} also called the MMEN construction into question. They argued that the total surface density (gas and dust, assuming a typical gas-to-dust ratio of 200 in the ISM) of the MMEN constructed from {\it Kepler} planets is close to the gravitational instability threshold \citep{Toomre}. \citet{Schlichting2014} then suggested that the majority of {\it Kepler} sub-Neptunes likely formed at $>2$AU before disk migration brought them in. However, the ALMA survey of Lupus by \citet{Ansdell} revealed that the enhancement of dusty materials in the inner disk is a ubiquitous phenomenon with the majority of systems showing gas-to-dust ratios closer to $10$ than $100$. Moreover, \citet{Schlichting2014} assumed a fixed temperature of 1000K throughout the whole disk. Here we use a more realistic temperature profile of $T\approx 150K (a/1{\rm AU})^{-0.6}$ \citep{Kenyon,Chiang}. After plugging into the Toomre stability criterion $Q\equiv c_s \Omega/\pi G \Sigma_{\rm tot}$, the critical total surface density is about $\Sigma_{\rm tot}=$1.2$\times10^{5}$ g cm$^{-2}$ $(a/1{\rm AU})^{-1.8}$. This is safely above the MMEN surface density (Eqn \ref{powerlaw_result} and \ref{RV_results}) with a reasonable choice of gas-to-dust ratio.

The analyses we presented in this work favored the MMEN idea particularly because the independent CKS and RV samples gave consistent results (Eqn \ref{powerlaw_result} and \ref{RV_results}). Furthermore, the concerns of MMEN pointed out by \citet{Schlichting2014} and \citet{Raymond} seem to be largely resolved in light of the new observational results \citep{Zhu2018,Dai,Ansdell}.

\begin{figure*}
\begin{center}
\includegraphics[width = 1.8\columnwidth]{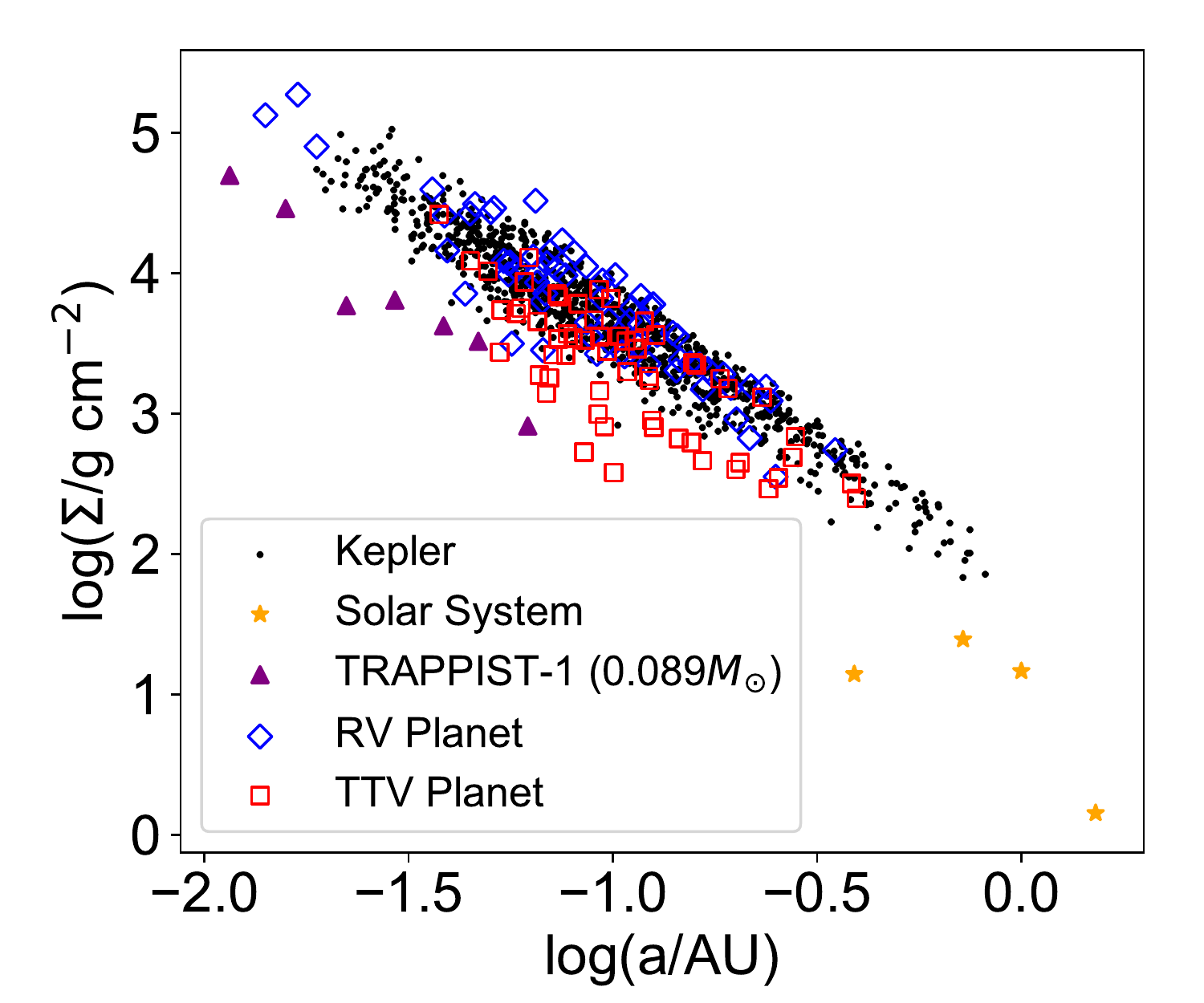}
\caption{ Same as Fig. \ref{a_vs_sigma}. In addition to the {\it Kepler} systems, we also plotted the planetary systems where the masses have been explicitly determined by radial velocity measurements (blue diamonds) and transit-timing variations (red Squares). We also included the TRAPPIST-1 system \citep[0.089$M_\odot$, purple triangle,][]{trappist} whose solid surface density is about one order of magnitude lower than sun-like stars  consistent with the expectation from Eqn. \ref{powerlaw_result}. The solar system terrestrial planets (yellow stars) also suggest a low solid surface density; we discuss this result further in Section \ref{sec:solar}.}
\label{fig:all_systems}
\end{center}
\end{figure*}

\subsection{The Solar System in the MMEN Context}
\label{sec:solar}
How should we understand the Solar System in the context of the MMEN? First of all, the MMSN \citep{Hayashi} is about an order of magnitude less dense than the MMEN (Eqn \ref{powerlaw_result}). If we include the four terrestrial planets of our Solar system in the MMEN analysis, they appear roughly one order of magnitude lower than the cloud of points from {\it Kepler} (see Fig. \ref{fig:all_systems}). Is our Solar system somehow unusual? Although the {\it Kepler} spacecraft achieved exquisite photometric precision, it is probably unable to discover the analogs of our terrestrial planets. This is demonstrated both in injection-recovery tests \citep{Christiansen} and in the large uncertainty of $\eta_{\oplus}$ calculation \citep[e.g.][]{Hsu}. The detection of our terrestrial planets is even more challenging in existing RV surveys. In short, the analogs of Solar system terrestrial planets are not well represented in the CKS or RV sample. The Solar system may be on the less massive end of the MMEN distribution that is yet to be explored observationally. However, our Solar system is not much of an outlier provided that only $\sim30$\% of sun-like stars host {\it Kepler}-like close-in super-Earths/sub-Neptunes \citep[1-4$R_\oplus$][]{Zhu2018}. The hosts of these {\it Kepler}-like systems are probably at the higher end of the solid surface density distribution. 

 It's also worth noting that the formation of our terrestrial planets might have been heavily influenced by the migration of Jupiter and Saturn in the Grand-Tack model \citep[e.g.][]{Walsh}. Such a migration history might not have happened for many {\it Kepler} systems. In addition, if pebble accretion \citep[e.g.][]{Lambrechts} was important, the {\it in-situ} formation narrative adopted by MMEN/MMSN analysis has to be revised in a future work.

\section{Summary}
In this work, we revisit the idea of MMEN specifically by investigating a possible correlation between MMEN solid surface density $\Sigma$ and host star properties $M_\star$ and [Fe/H]. This work offers a fresh perspective on the formation of {\it Kepler}-like sub-Neptune planets ($1R_\oplus<R_p<4R_\oplus$; $a$<1AU) across spectral type and metallicity. In comparison with the more traditional occurrence rate study, the MMEN framework incorporates more information such as the multiplicity of planets, the orbital spacing and the planet size correlation within a system.

\begin{itemize}
    \item {We constructed the MMEN using the uniform and highly precise CKS sample \citep{CKS1}. We converted the transit radii to planetary mass using a set of mass-radius relationships reported in the literature. We also experimented with different assumptions of the planet feeding zones.}

    \item {We modeled the correlation of MMEN solid surface density $\Sigma$, orbital distance $a$, host star mass $M_\star$ and metallicity [Fe/H] with a simple power law model. We performed model selection with Bayesian Evidence calculation and sampled parameter posterior distribution with the nested-sampling code {\sc MultiNest}.}

    \item {We found a strong, almost linear correlation between $\Sigma$ and $M_\star$ and a weak, sub-linear correlation between $\Sigma$ and [Fe/H] (Eqn \ref{powerlaw_result}). This shows that the formation of sub-Neptune planets proceeds readily in lower-metallicity environments while giant planet formation strongly favors metal-rich systems. Meanwhile, processes that would weaken $\Sigma$-[Fe/H] correlation also play a part including radial drift of dust, emergence of giant planets, dynamical instability etc.}
    
    \item{The RV and TTV sample eliminated the need for mass-radius relationships. It is reassuring for the {\it in-situ} formation theories that CKS and RV samples produced consistent results for the MMEN (Eqn \ref{powerlaw_result} and \ref{RV_results}). On other hand, the MMEN constructed from TTV planets shows weaker dependence on host star properties. This may be related to the convergent migration undergone by the TTV systems breaking the {\it in-situ} assumption of MMEN.  Convergent migration of TTV planet has been previously suggested \citep{Lee,MillsNature}.}

    \item {The occurrence rate of sub-Neptune planets declines rapidly for $a<$0.1AU. However, MMEN does not show a corresponding drop in solid surface density $\Sigma$. This favors a ``Solid Disk Truncation'' picture \citep[magnetosphere truncation][]{Lee_usp} rather than a ``Solid Disk Thinning'' picture (dust sublimation or dust radial drift).}
    
    \item {Comparing {\it Kepler} single-transiting and multi-transiting systems, we found that the singles show a much weaker correlation of $\Sigma$-[Fe/H] and $\Sigma$-$M_\star$. This may be ascribed to the dynamically hot evolution (giant impacts and interaction with cold Jupiters) that tend to reduce the planet multiplicity and induce larger mutual inclinations. The dynamically hot evolution tends to produce single-transiting systems and also modifies the initial orbital architecture from formation in a disk.}
    
    \item {Lower-mass stars seem to have a higher efficiency of forming/retaining planets.  About 22\% of the solid mass within $\sim$1AU of the disks around sun-like stars are converted/preserved as planets. On the other hand, the number goes up to about 67\% for M stars. The reason may be a combination of the lower binary fraction, lower giant planet occurrence rate and the longer disk lifetime of lower-mass stars.}
    
    \item{MMEN $\Sigma$ shows a $\sim$0.2 dex intrinsic variation that cannot be attributed to measurement uncertainties. This scatter attests the variability in disk properties as well as the stochasticity of planet formation.}

\end{itemize}

\software{MultiNest \citep{Feroz}, Forecaster \citep{Chen2017}}

\acknowledgements
We thank Heather Knutson, Kento Masuda, Luke Bouma, Sarah Millholland, Sharon Wang, Ji-Wei Xie, Doug Lin and Eve Lee for helpful discussions.

\bibliography{main}

\appendix{}
\renewcommand{\thefigure}{A\arabic{figure}}

\setcounter{figure}{0}

\begin{figure*}
\begin{center}
\includegraphics[width = 0.3\columnwidth]{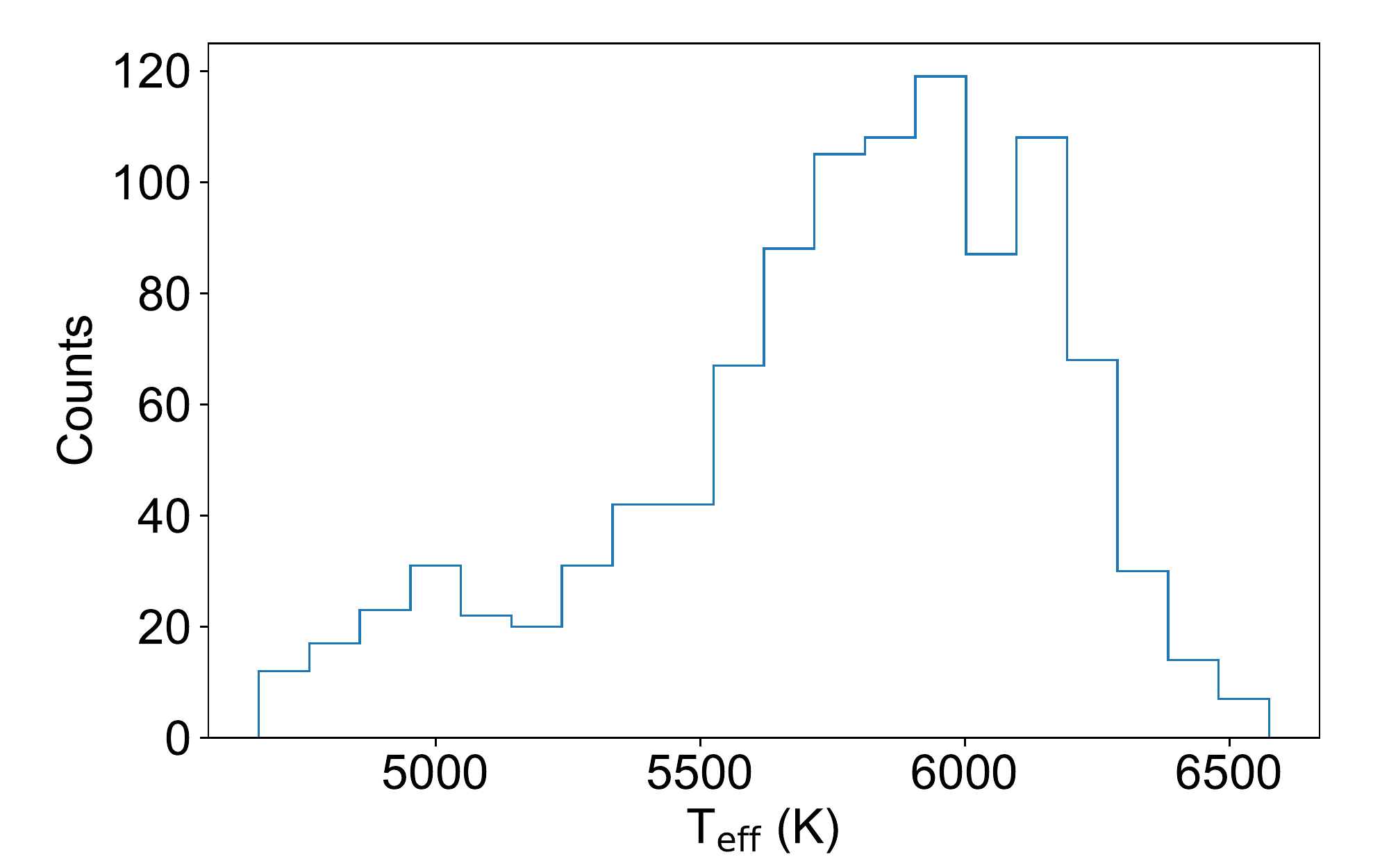}
\includegraphics[width = .3\columnwidth]{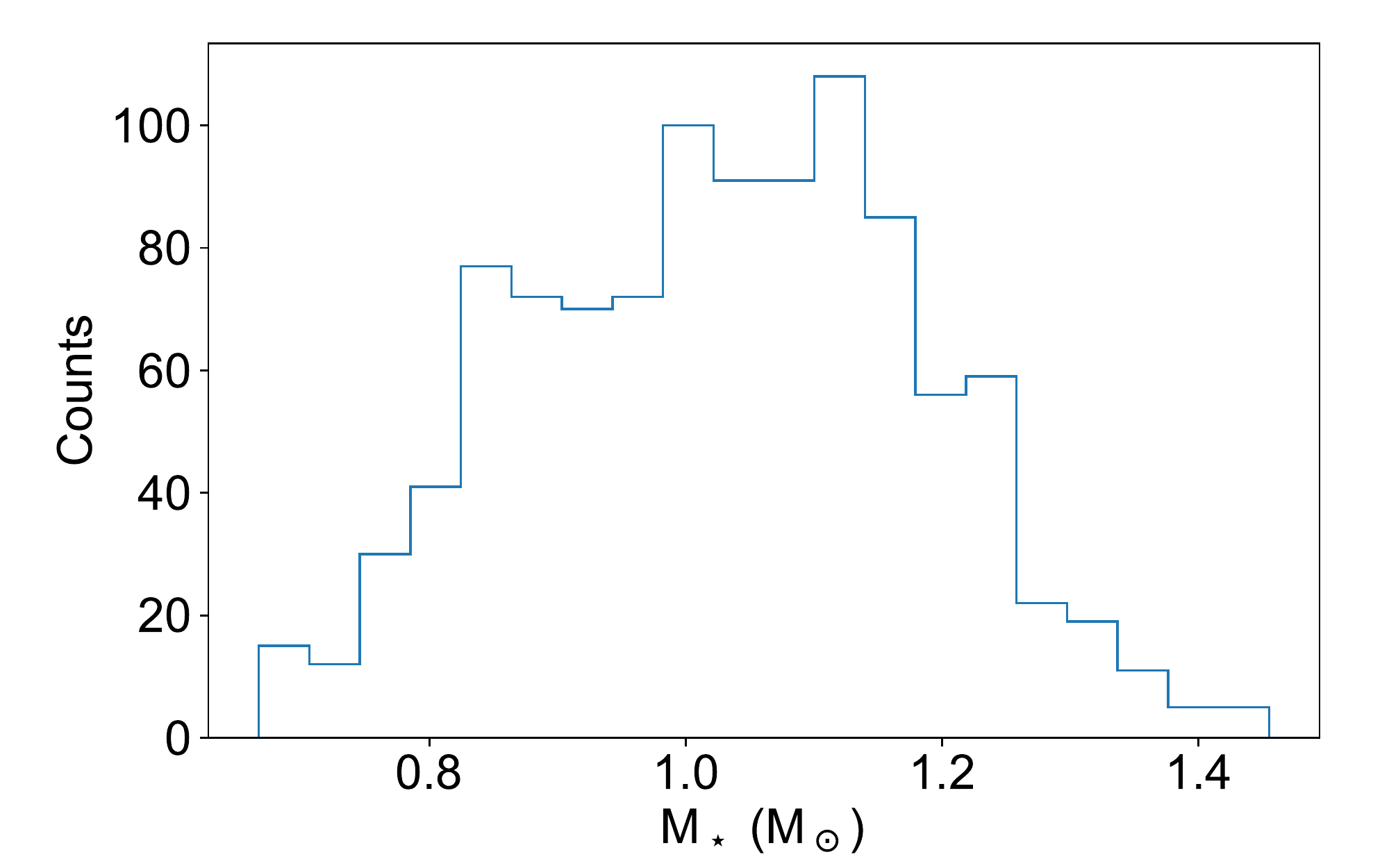}
\includegraphics[width = .3\columnwidth]{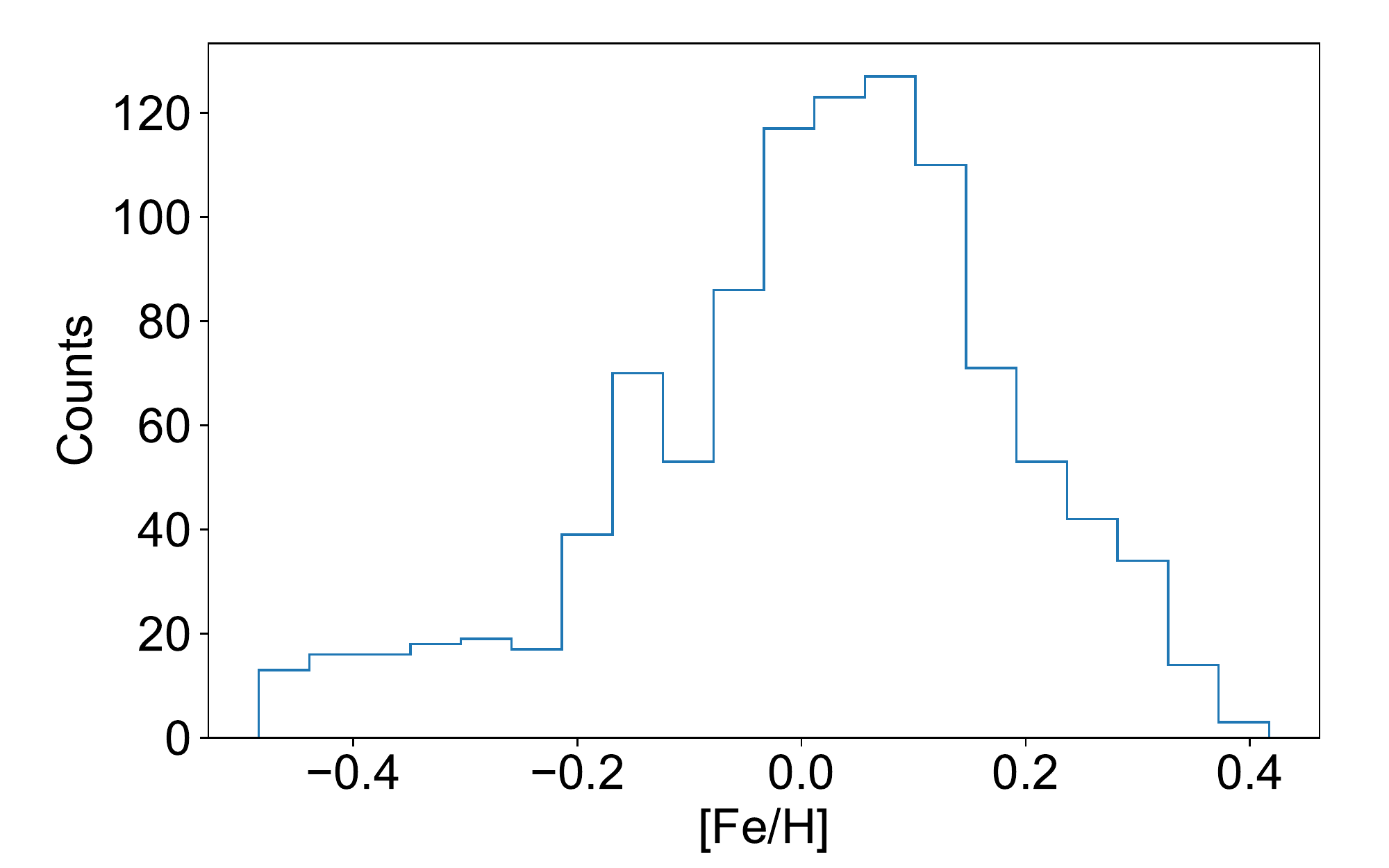}
\includegraphics[width = .3\columnwidth]{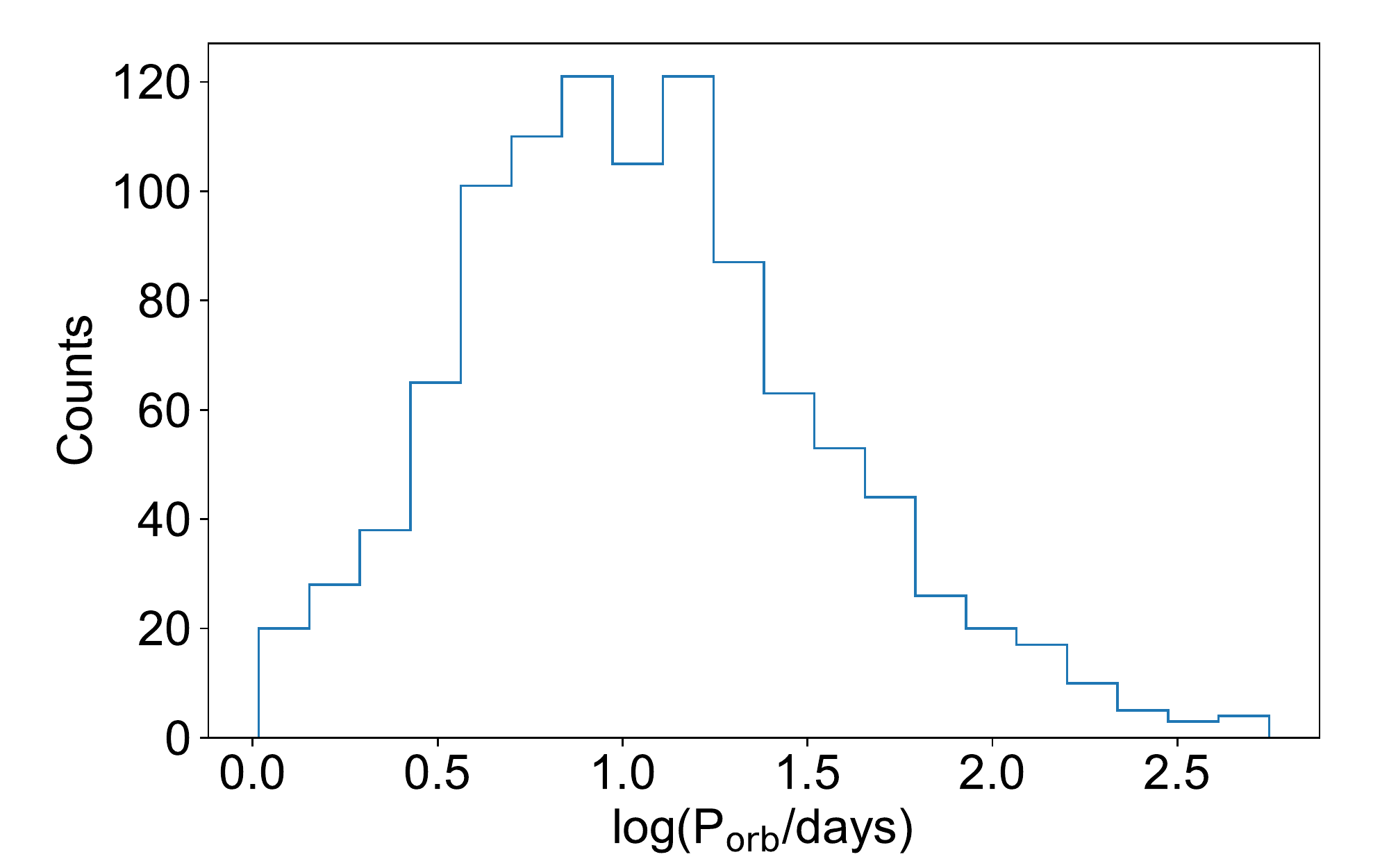}
\includegraphics[width = .3\columnwidth]{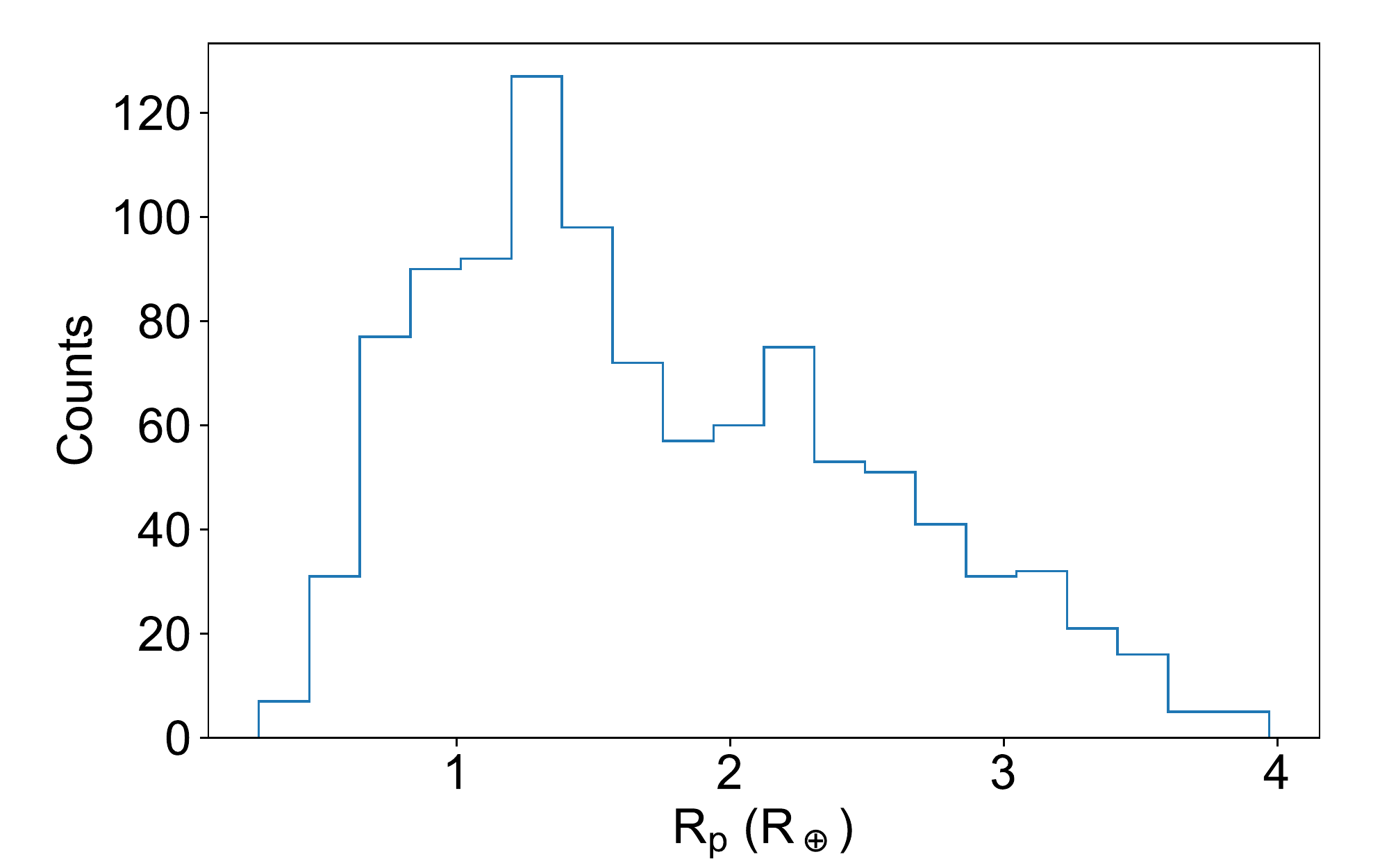}
\includegraphics[width = .3\columnwidth]{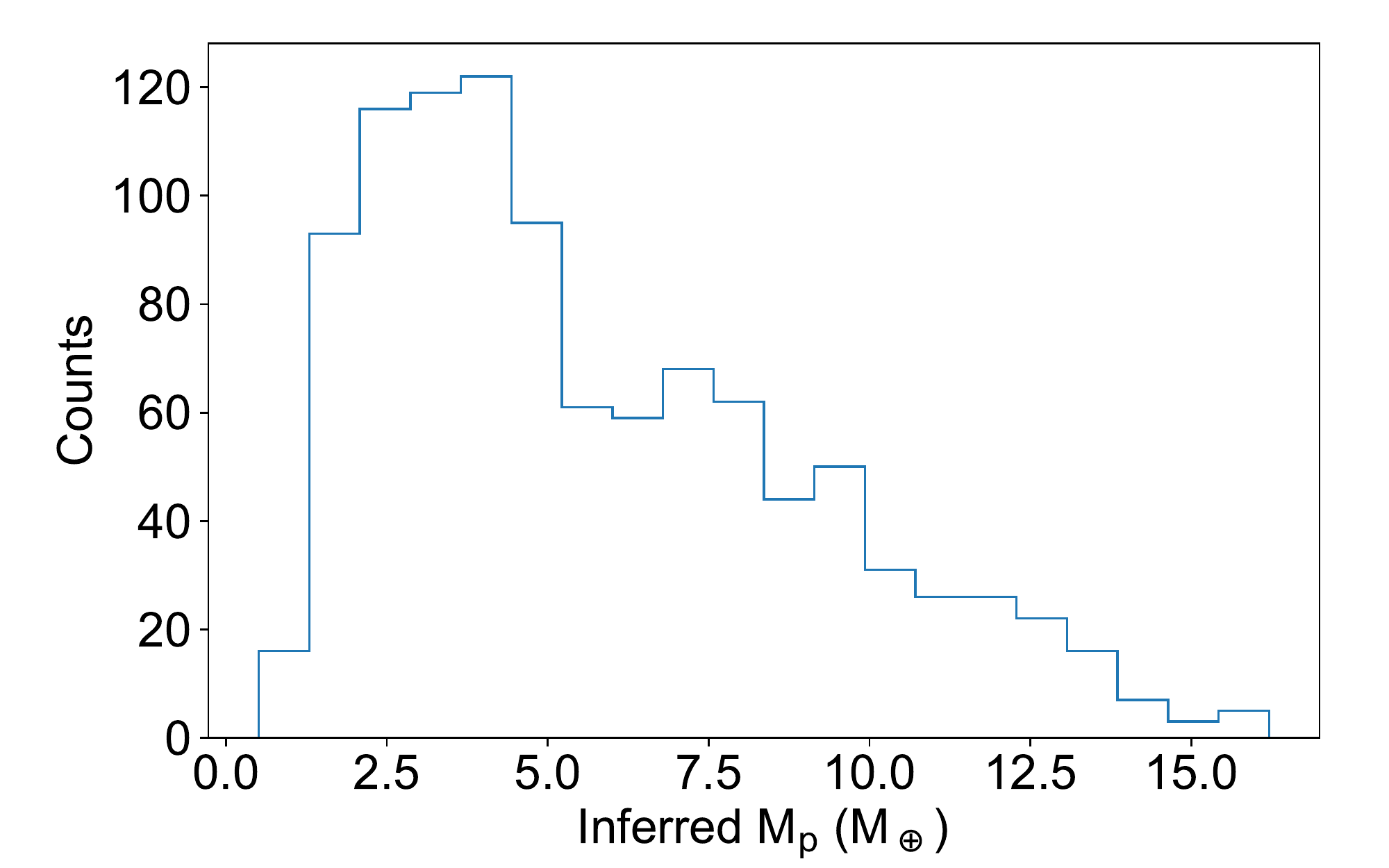}
\caption{The CKS sample: the histograms of the stellar effective temperature $T_{\rm eff}$, stellar mass $M_\star$, stellar metallicity [Fe/H], the orbital period $P_{\rm orb}$, the planetary radius $R_{\rm p}$ and the planetary mass $M_p$ inferred from the mass-radius relationship of \citet{Wolfgang2016}.}
\label{the_sample}
\end{center}
\end{figure*}

\begin{figure*}
\begin{center}
\includegraphics[width = 1.\columnwidth]{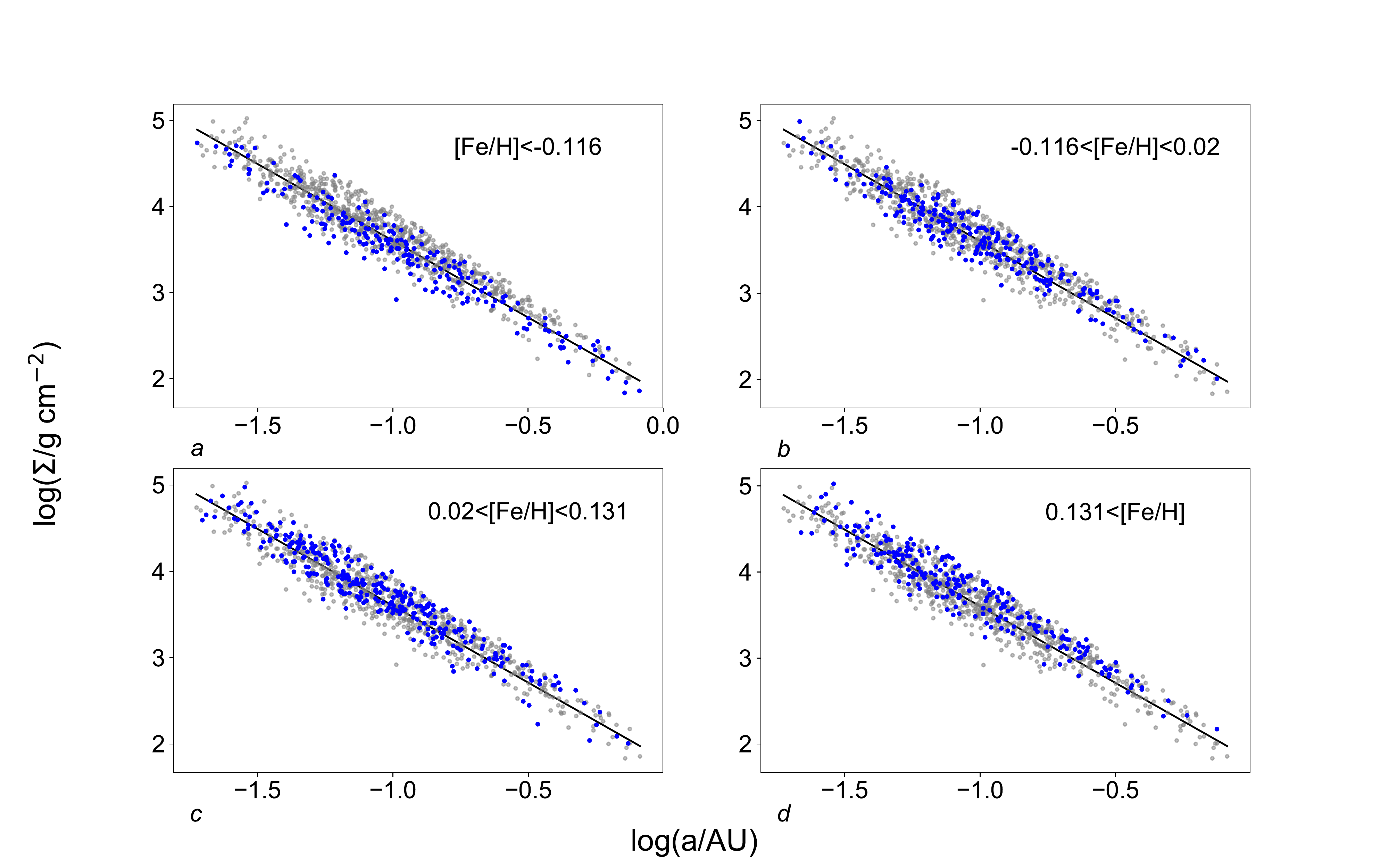}
\includegraphics[width = 1.\columnwidth]{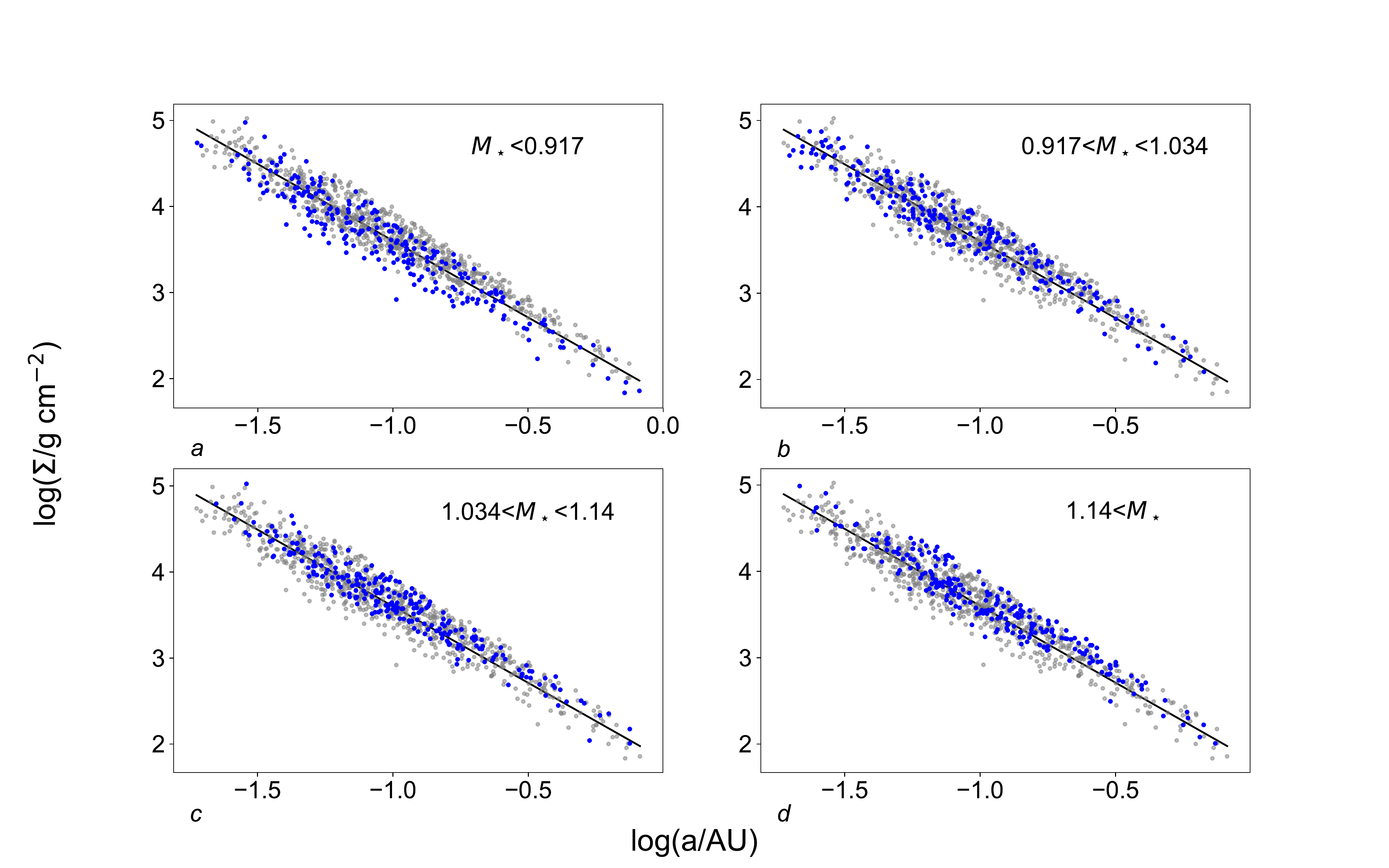}
\caption{The same as Fig. \ref{a_vs_sigma} but split into four quartiles of stellar metallcity [Fe/H] and stellar mass $M_\star$. More massive and more metal-rich systems systematically have higher solid surface density.}
\label{four_panel}
\end{center}
\end{figure*}

\begin{figure}
\begin{center}
\includegraphics[width = .75\columnwidth]{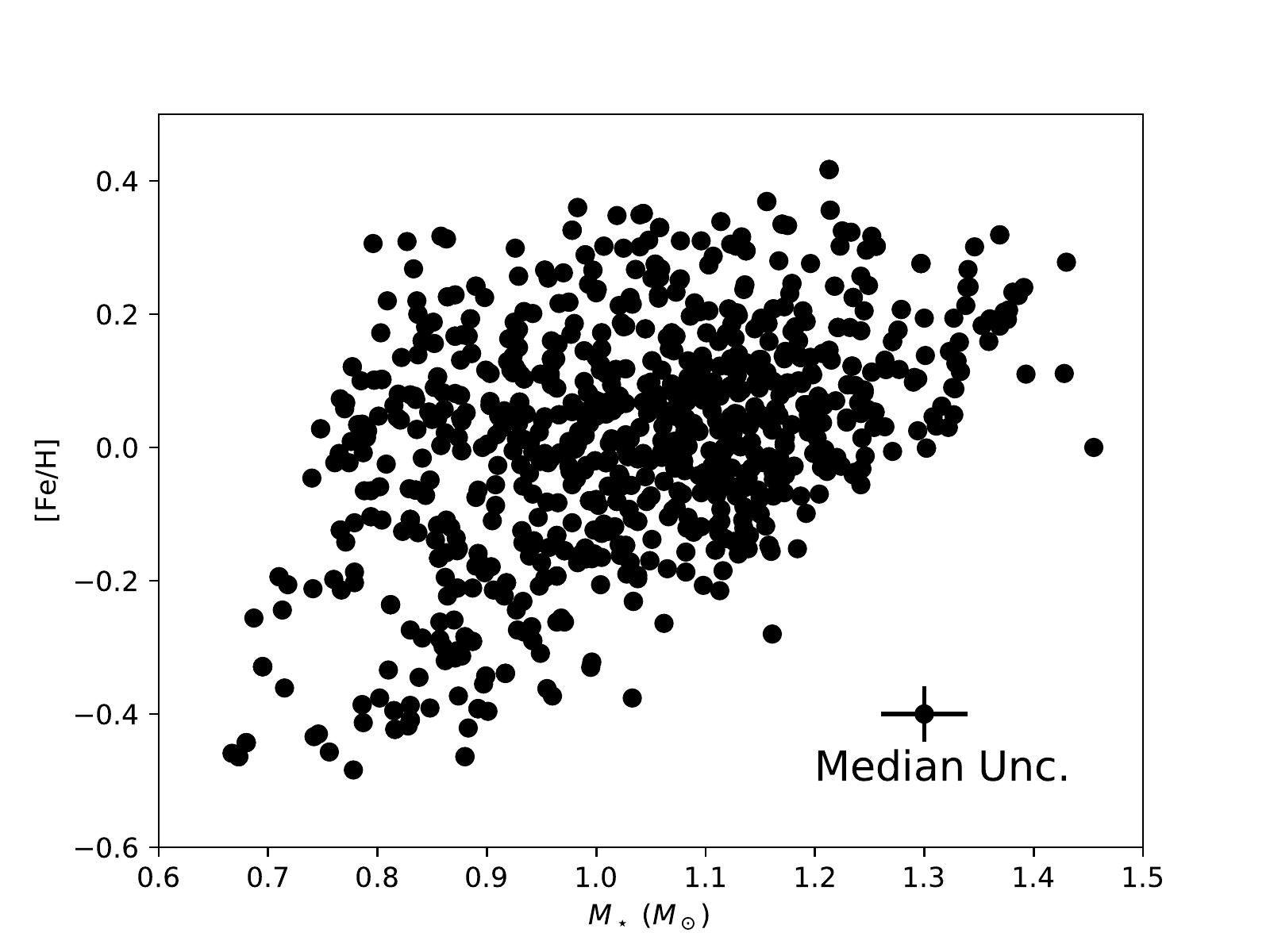}
\caption{The stellar mass and metallicity of the stars in CKS sample shows strong positive correlation. This correlation may be resulted from Galactic chemical evolution and a sample selection effect in Fig.\ref{correlation}.}
\label{ms_vs_met}
\end{center}
\end{figure}

\begin{figure*}
\begin{center}
\includegraphics[width = .75\columnwidth]{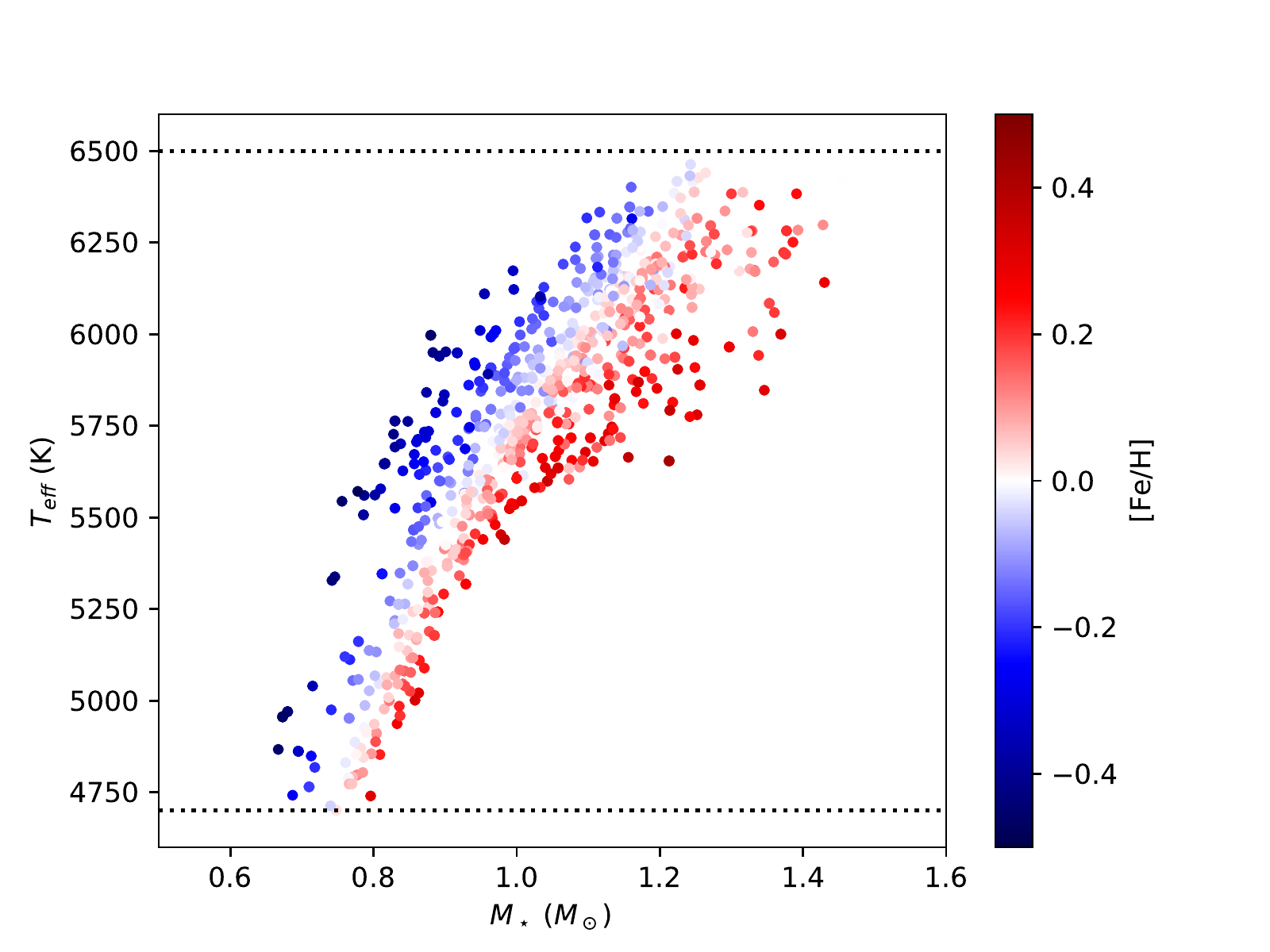}
\includegraphics[width = .75\columnwidth]{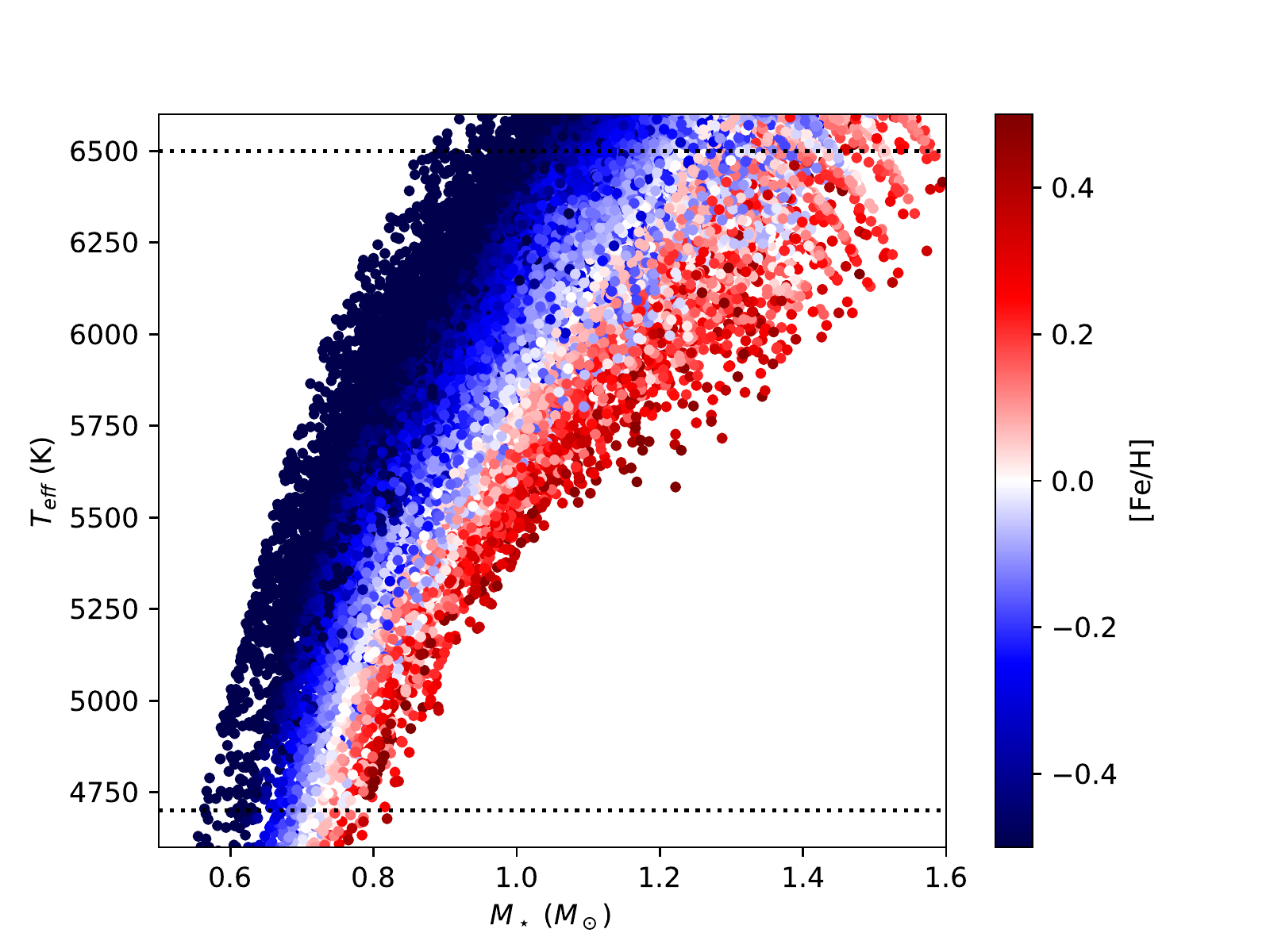}
\caption{The scatter plots of stellar effective temperature $T_{\rm eff}$ versus stellar mass $M_\star$ for the CKS stars and Kepler Input Catalog (KIC) stars \citep[stellar properties from][]{Mathur}. The color coding is by stellar metallicity [Fe/H]. At the same stellar mass, a lower host star metallicity gives a lower opacity in the stellar atmosphere. The star thus has smaller radii but higher effective temperature. Because the CKS sample was selected by slicing in the effective temperature (4700K-6500K), the sample excluded higher mass stars ($\gtrsim$ 1.2$M_\star$) with sub-solar metallicity ([Fe/H]<0). There is a analogous but opposite effect for the lower mass end. This effect partially accounts for the $M_\star$-[Fe/H] covariance seen in Fig. \ref{ms_vs_met}.}
\label{correlation}
\end{center}
\end{figure*}

\begin{figure*}
\begin{center}
\includegraphics[width = .75\columnwidth]{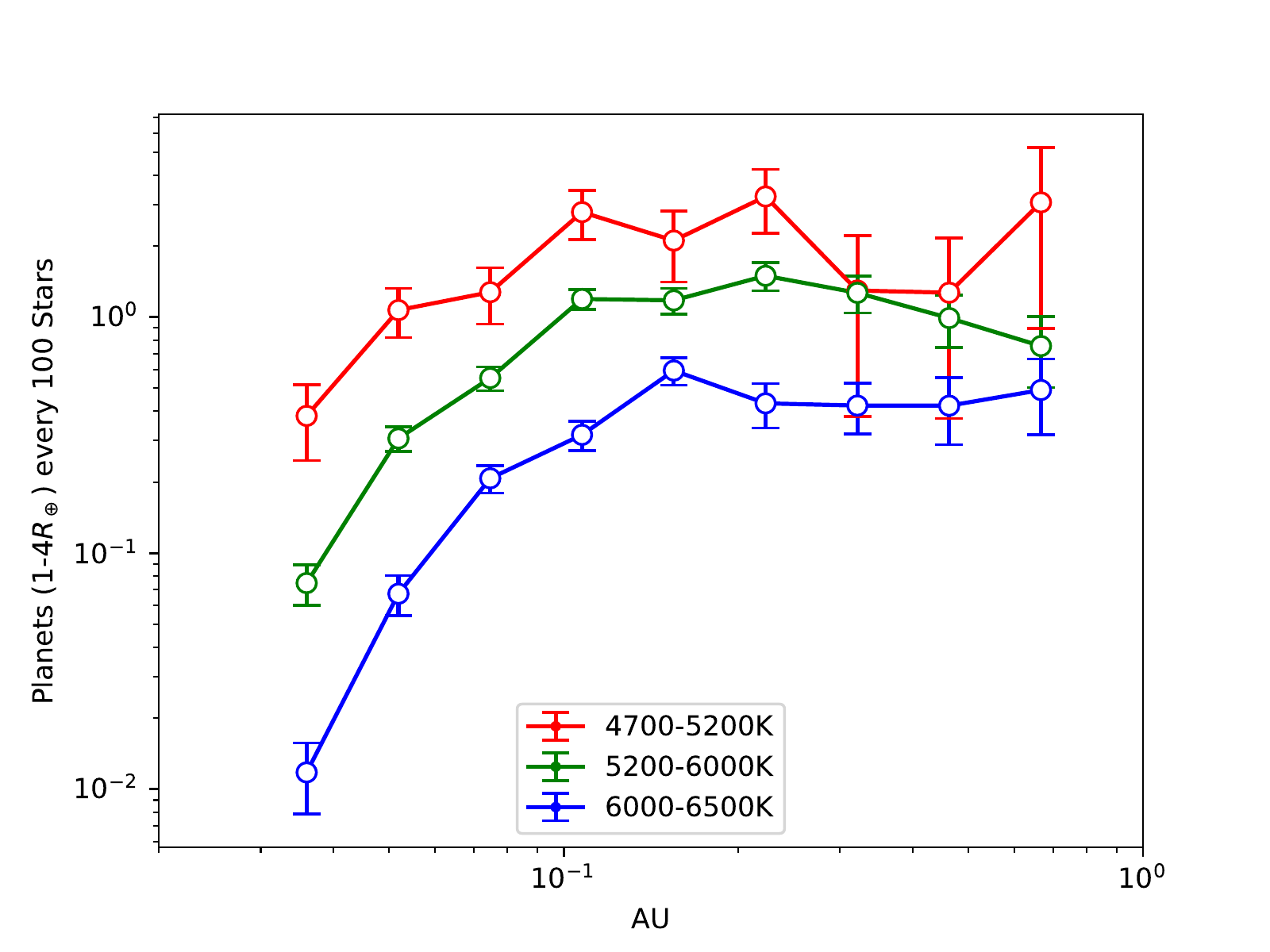}
\caption{The occurrence rate of sub-Neptune planets (1-4 $R_\oplus$) as a function of orbital distance. We calculated separately the occurrence rates for different bins of host star effective temperature. We are able to reproduce the previous claims that 1) lower-mass stars are more likely to host sub-Neptune planets \citep[e.g.][]{Mulders2015,Dressing}; 2) regardless of host star spectral types, the occurrence of sub-Neptune planets show a broken power where the occurrence rate increases steadily with orbital distance until $a\approx0.1$AU after which the occurrence rates plateau out to about 1AU.}
\label{occurrence}
\end{center}
\end{figure*}

\begin{deluxetable*}{llllllllll}
\tablecaption{Summary of the Power Law Models (Eqn \ref{eqn:powerlaw}) for the CKS and KOI Sample \label{tab:powerlaw}}
\tablehead{
\colhead{Mass-Radius Relation} & \colhead{log($\Sigma_0$/g cm$^{-3}$)} & \colhead{$l$ ($a$ index)} & \colhead{$m$ ([Fe/H] index)} & \colhead{$n$ ($M_\star$ index)} & \colhead{Scatter $\Delta$log($\Sigma_0$)}  & \colhead{Evidence $\Delta$log($Z$)}}
\startdata
CKS\\
\citet{Lissauer2011} &  1.812 $\pm$ 0.022 & -1.609$\pm$ 0.025 & - & - & 0.25 & -19.8 \\
\citet{Lissauer2011} &  1.840 $\pm$ 0.021 & -1.577$\pm$ 0.025 & 0.469 $\pm$  0.044 &-  & 0.23 & -2.5\\
\citet{Lissauer2011} &  1.793 $\pm$ 0.022 & -1.630$\pm$ 0.025 & - & 1.12 $\pm$ 0.11 & 0.23 & -1.8\\
\citet{Lissauer2011} &  1.821 $\pm$ 0.022 & -1.599$\pm$ 0.024 & 0.336 $\pm$  0.051 & 0.683 $\pm$ 0.13 & 0.23 & 0\\
\\
\citet{Weiss2014} & 1.826 $\pm$ 0.024 & -1.615$\pm$ 0.027 & -  & -  & 0.26 & -17.6\\
\citet{Weiss2014} & 1.852 $\pm$ 0.024 & -1.585$\pm$ 0.026 & 0.424 $\pm$ 0.049 & -  & 0.26 & -7.2\\
\citet{Weiss2014} & 1.807 $\pm$ 0.022 & -1.636$\pm$ 0.025 & - & 1.18 $\pm$ 0.12 & 0.25 & 0\\
\citet{Weiss2014} & 1.829 $\pm$ 0.024 & -1.610$\pm$ 0.026 & 0.265 $\pm$ 0.055 & 0.84 $\pm$ 0.14 & 0.25 & -1.9 \\
\\
\citet{Wolfgang2016} &  1.864 $\pm$ 0.014 & -1.751$\pm$ 0.015 & - & - & 0.11 & -23.4\\
\citet{Wolfgang2016} &  1.883 $\pm$ 0.014 & -1.729$\pm$ 0.016 & 0.317 $\pm$ 0.028 & - & 0.10 & -12.9\\
\citet{Wolfgang2016} &  1.850 $\pm$ 0.013 & -1.767$\pm$ 0.015 & - & 0.83 $\pm$ 0.07 & 0.10 & 0\\
\citet{Wolfgang2016} &  1.867 $\pm$ 0.014 & -1.747$\pm$ 0.016 & 0.212 $\pm$ 0.033 & 0.56 $\pm$ 0.09 & 0.10 & -7.4\\
\\
\citet{Chen2017} &  1.700 $\pm$ 0.018 & -1.677$\pm$ 0.021 & - & -& 0.20 & -18.7\\
\citet{Chen2017} &  1.723 $\pm$ 0.018 & -1.649$\pm$ 0.020 & 0.394 $\pm$ 0.037 & - & 0.19 & -4.3\\
\citet{Chen2017} &  1.684 $\pm$ 0.018 & -1.695$\pm$ 0.019 & -  & 0.99 $\pm$ 0.10& 0.19 & 0\\
\citet{Chen2017} &  1.705 $\pm$ 0.018 & -1.669$\pm$ 0.021 & 0.278 $\pm$ 0.042 & 0.619 $\pm$ 0.11& 0.19 & -1.2\\
\\
\citet{Mills} &  1.781 $\pm$ 0.010 & -1.808$\pm$ 0.012 & - & - & 0.12 & -28.3\\
\citet{Mills} &  1.796 $\pm$ 0.011 & -1.791$\pm$ 0.012 & 0.257 $\pm$  0.022 & - & 0.12 & -24.3\\
\citet{Mills} &  1.768 $\pm$ 0.010 & -1.822$\pm$ 0.011 & - & 0.713 $\pm$ 0.052 & 0.11 & 0\\
\citet{Mills} &  1.781 $\pm$ 0.011 & -1.808$\pm$ 0.012 & 0.162 $\pm$  0.025 & 0.50 $\pm$ 0.06 & 0.11 & -17.2\\
\hline
KOI\\
\citet{Lissauer2011} &  1.866 $\pm$ 0.027 & -1.586$\pm$ 0.031 & - & - & 0.26 & 0\\
\citet{Lissauer2011} &  1.876 $\pm$ 0.028 & -1.572$\pm$ 0.031 & 0.44 $\pm$  0.10 &-  & 0.26 & -1.1\\
\citet{Lissauer2011} &  1.849 $\pm$ 0.028 & -1.592$\pm$ 0.031 & - & 0.56 $\pm$ 0.29 & 0.25 & -0.9\\
\citet{Lissauer2011} &  1.860 $\pm$ 0.029 & -1.578$\pm$ 0.031 & 0.42 $\pm$  0.10 & 0.45 $\pm$ 0.30 & 0.26 & -2.3\\
\\
\citet{Weiss2014} & 1.855 $\pm$ 0.029 & -1.627$\pm$ 0.032 & -  & -  & 0.22 & 0\\
\citet{Weiss2014} & 1.860 $\pm$ 0.027 & -1.617$\pm$ 0.031 & 0.32 $\pm$ 0.10 & -  & 0.22 & -3.4\\
\citet{Weiss2014} & 1.830 $\pm$ 0.030 & -1.634$\pm$ 0.031 & - & 0.83 $\pm$ 0.30 & 0.22 & -0.1\\
\citet{Weiss2014} & 1.836 $\pm$ 0.030 & -1.627$\pm$ 0.032 & 0.29 $\pm$ 0.10 & 0.77 $\pm$ 0.32 & 0.22 & -3.4\\
\\
\citet{Wolfgang2016} &  1.900 $\pm$ 0.018 & -1.737$\pm$ 0.020 & - & - & 0.08 & 0\\
\citet{Wolfgang2016} &  1.906 $\pm$ 0.019 & -1.729$\pm$ 0.020 & 0.281 $\pm$ 0.068 & - & 0.09 & -6.6\\
\citet{Wolfgang2016} &  1.885 $\pm$ 0.019 & -1.742$\pm$ 0.020 & - & 0.48 $\pm$ 0.19 & 0.08 & -1.2\\
\citet{Wolfgang2016} &  1.892 $\pm$ 0.018 & -1.733$\pm$ 0.020 & 0.269 $\pm$ 0.065 & 0.41 $\pm$ 0.19 & 0.09 & -8.0\\
\\
\citet{Chen2017} &  1.744 $\pm$ 0.023 & -1.660$\pm$ 0.027 & - & -& 0.18 & 0\\
\citet{Chen2017} &  1.751 $\pm$ 0.023 & -1.648$\pm$ 0.026 & 0.361 $\pm$ 0.082 & - & 0.18 & -2.9 \\
\citet{Chen2017} &  1.729 $\pm$ 0.024 & -1.664$\pm$ 0.025 & -  & 0.53 $\pm$ 0.26& 0.18 & -0.9\\
\citet{Chen2017} &  1.738 $\pm$ 0.025 & -1.653$\pm$ 0.026 & 0.350 $\pm$ 0.086 & 0.43 $\pm$ 0.25& 0.18 & -4.1 \\
\\
\citet{Mills} &  1.808 $\pm$ 0.014 & -1.800$\pm$ 0.015 & - & - & 0.10 & 0\\
\citet{Mills} &  1.812 $\pm$ 0.014 & -1.793$\pm$ 0.016 & 0.213 $\pm$  0.052 & - & 0.11 & -12.2\\
\citet{Mills} &  1.795 $\pm$ 0.014 & -1.804$\pm$ 0.016 & - & 0.45 $\pm$ 0.14 & 0.10 & -1.7\\
\citet{Mills} &  1.800 $\pm$ 0.015 & -1.797$\pm$ 0.017 & 0.203 $\pm$  0.050 & 0.39 $\pm$ 0.15 & 0.11 & -14\\
\hline
\enddata
\end{deluxetable*}

\begin{deluxetable*}{llllllllll}
\tablecaption{Summary of the Power Law Models for the CKS Single-transiting and Multi-transiting Systems\label{tab:single_multi}}
\tablehead{
\colhead{Mass-Radius Relation} & \colhead{log($\Sigma_0$/g cm$^{-3}$)} & \colhead{$l$ ($a$ index)} & \colhead{$m$ ([Fe/H] index)} & \colhead{$n$ ($M_\star$ index)} & \colhead{Scatter $\Delta$log($\Sigma_0$)}  & \colhead{Evidence $\Delta$log($Z$)}}
\startdata
CKS Single-Transiting\\
\citet{Lissauer2011} &  1.805 $\pm$ 0.023 & -1.628$\pm$ 0.029 & - & - & 0.21 & -0.6\\
\citet{Lissauer2011} &  1.841 $\pm$ 0.024 & -1.592$\pm$ 0.030 & 0.340 $\pm$  0.069 &-  & 0.21 & 0\\
\citet{Lissauer2011} &  1.809 $\pm$ 0.023 & -1.626$\pm$ 0.029 & - & 0.51 $\pm$ 0.16 & 0.21 & -0.2\\
\citet{Lissauer2011} &  1.838 $\pm$ 0.024 & -1.595$\pm$ 0.030 & 0.31 $\pm$  0.07 & 0.23 $\pm$ 0.16 & 0.21 & -1.4\\
\\
\citet{Weiss2014} & 1.814 $\pm$ 0.022 & -1.640$\pm$ 0.028 & -  & -  & 0.20 & -0.6\\
\citet{Weiss2014} & 1.840 $\pm$ 0.023 & -1.614$\pm$ 0.028 & 0.251 $\pm$ 0.063 & -  & 0.20 & -2.6\\
\citet{Weiss2014} & 1.816 $\pm$ 0.023 & -1.640$\pm$ 0.028 & - & 0.52 $\pm$ 0.15 & 0.20 & 0\\
\citet{Weiss2014} & 1.836 $\pm$ 0.023 & -1.620$\pm$ 0.028 & 0.20 $\pm$ 0.07 & 0.35 $\pm$ 0.16 & 0.20 & -3.5\\
\\
\citet{Wolfgang2016} &  1.857 $\pm$ 0.015 & -1.765$\pm$ 0.020 & - & - & 0.07 & -1.4\\
\citet{Wolfgang2016} &  1.882 $\pm$ 0.016 & -1.741$\pm$ 0.020 & 0.24 $\pm$  0.04 & - & 0.07 & -4.9\\
\citet{Wolfgang2016} &  1.859 $\pm$ 0.015 & -1.765$\pm$ 0.018 & - & 0.45 $\pm$ 0.10 & 0.06 & 0\\
\citet{Wolfgang2016} &  1.878 $\pm$ 0.016 & -1.745$\pm$ 0.020 & 0.19 $\pm$  0.05 & 0.27 $\pm$ 0.11 & 0.07 & -6.1\\
\\
\citet{Chen2017} &  1.692 $\pm$ 0.019 & -1.694$\pm$ 0.024 & - & -& 0.17 & -0.8\\
\citet{Chen2017} &  1.722 $\pm$ 0.020 & -1.663$\pm$ 0.025 & 0.293 $\pm$ 0.055 & - & 0.17 & -1.3\\
\citet{Chen2017} &  1.695 $\pm$ 0.020 & -1.692$\pm$ 0.025 & -  & 0.48 $\pm$ 0.13& 0.17 & 0\\
\citet{Chen2017} &  1.720 $\pm$ 0.020 & -1.667$\pm$ 0.025 & 0.251 $\pm$ 0.060 & 0.25 $\pm$ 0.14& 0.17 & -2.8\\
\\
\citet{Mills} &  1.774 $\pm$ 0.012 & -1.821$\pm$ 0.015 & - & - & 0.10 & -2.6\\
\citet{Mills} &  1.794 $\pm$ 0.012 & -1.801$\pm$ 0.015 & 0.193 $\pm$  0.034 & - & 0.10 & -10\\
\citet{Mills} &  1.775 $\pm$ 0.012 & -1.821$\pm$ 0.014 & - & 0.421 $\pm$ 0.080 & 0.10 & 0\\
\citet{Mills} &  1.790 $\pm$ 0.012 & -1.805$\pm$ 0.016 & 0.148 $\pm$  0.038 & 0.28 $\pm$ 0.10 & 0.10 & -11.3\\
\hline
CKS Multi-Transiting\\
\citet{Lissauer2011} &  1.677 $\pm$ 0.033 & -1.703$\pm$ 0.040 & - & - & 0.29 & -17\\
\citet{Lissauer2011} &  1.705 $\pm$ 0.033 & -1.670$\pm$ 0.038 & 0.532 $\pm$  0.062 &-  & 0.28 & -6.1\\
\citet{Lissauer2011} &  1.680 $\pm$ 0.031 & -1.708$\pm$ 0.037 & - & 1.67 $\pm$ 0.18 & 0.27 & 0\\
\citet{Lissauer2011} &  1.695 $\pm$ 0.031 & -1.689$\pm$ 0.038 & 0.297 $\pm$  0.071 & 1.19 $\pm$ 0.21 & 0.27 & -0.6\\
\\
\citet{Weiss2014} & 1.702 $\pm$ 0.039 & -1.684$\pm$ 0.045 & -  & -  & 0.34 & -13.4\\
\citet{Weiss2014} & 1.727 $\pm$ 0.039 & -1.657$\pm$ 0.046 & 0.481 $\pm$ 0.077 & -  & 0.33 & -7.7\\
\citet{Weiss2014} & 1.707 $\pm$ 0.038 & -1.692$\pm$ 0.044 & - & 1.71 $\pm$ 0.21 & 0.32 & 0\\
\citet{Weiss2014} & 1.714 $\pm$ 0.037 & -1.677$\pm$ 0.044 & 0.209 $\pm$ 0.087 & 1.37 $\pm$ 0.25 & 0.32 & -2.3\\
\\
\citet{Wolfgang2016} &  1.777 $\pm$ 0.022 & -1.811$\pm$ 0.027 & - & - & 0.15 & -20.6 \\
\citet{Wolfgang2016} &  1.795 $\pm$ 0.020 & -1.790$\pm$ 0.025 & 0.36 $\pm$  0.04 & - & 0.14 & -11.6\\
\citet{Wolfgang2016} &  1.779 $\pm$ 0.020 & -1.816$\pm$ 0.024 & - & 1.16 $\pm$ 0.11 & 0.13 & 0\\
\citet{Wolfgang2016} &  1.788 $\pm$ 0.020 & -1.802$\pm$ 0.024 & 0.19 $\pm$  0.05 & 0.87 $\pm$ 0.13 & 0.15 & -4.7\\
\\
\citet{Chen2017} &  1.586 $\pm$ 0.028 & -1.756$\pm$ 0.033 & - & -& 0.24 & -18.2\\
\citet{Chen2017} &  1.610 $\pm$ 0.027 & -1.727$\pm$ 0.032 & 0.449 $\pm$ 0.054 & - & 0.23 & -7.9\\
\citet{Chen2017} &  1.590 $\pm$ 0.026 & -1.758$\pm$ 0.032 & -  & 1.42 $\pm$ 0.14& 0.22 & 0\\
\citet{Chen2017} &  1.601 $\pm$ 0.025 & -1.744$\pm$ 0.030 & 0.245 $\pm$ 0.059 & 1.04 $\pm$ 0.17& 0.22 & -2.0\\
\\
\citet{Mills} &  1.712 $\pm$ 0.016 & -1.856$\pm$ 0.019 & - & - & 0.14 & -23.4\\
\citet{Mills} &  1.728 $\pm$ 0.016 & -1.839$\pm$ 0.020 & 0.288 $\pm$  0.031 & - & 0.14 & -17\\
\citet{Mills} &  1.714 $\pm$ 0.015 & -1.860$\pm$ 0.018 & - & 0.97 $\pm$ 0.08 & 0.13 & 0\\
\citet{Mills} &  1.722 $\pm$ 0.016 & -1.850$\pm$ 0.018 & 0.143 $\pm$  0.036 & 0.74 $\pm$ 0.10 & 0.13 & -9\\
\enddata
\end{deluxetable*}

\begin{deluxetable*}{llllllllll}
\tablecaption{Summary of the Power Law Models for Planetary Systems with Mass Measurements from RV and TTV \label{tab:ttv_rv}}
\tablehead{
\colhead{} & \colhead{log($\Sigma_0$/g~cm$^{-3}$)} & \colhead{$l$ ($a$ index)} & \colhead{$m$ ([Fe/H] index)} & \colhead{$n$ ($M_\star$ index)} & \colhead{Scatter $\Delta$log($\Sigma_0$)}  & \colhead{Evidence $\Delta$log($Z$)}}
\startdata
TTV Systems\\
- &  1.860 $\pm$ 0.081 & -1.689$\pm$ 0.087 & - & - & 0.15 & 0\\
- &  1.850 $\pm$ 0.082 & -1.699$\pm$ 0.087 & 0.09 $\pm$  0.16 & - & 0.15 & -2.\\
- &  1.818 $\pm$ 0.085 & -1.76$\pm$ 0.10 & - & 0.38 $\pm$ 0.24 & 0.14 & -0.8\\
- &  1.814 $\pm$ 0.091 & -1.76$\pm$ 0.10 & 0.02 $\pm$  0.17 & 0.38 $\pm$ 0.25 & 0.15 & -2.8\\
\hline
RV Systems\\
- &  1.98 $\pm$ 0.10 & -1.74$\pm$ 0.11 & - & - & 0.19 & -2.9\\
- &  2.00 $\pm$ 0.11 & -1.72$\pm$ 0.12 & 0.29 $\pm$  0.17 & - & 0.21 & -6.2\\
- &  1.99 $\pm$ 0.09 & -1.78$\pm$ 0.10 & - & 0.89 $\pm$ 0.19 & 0.16 & 0\\
- &  1.99 $\pm$ 0.10 & -1.77$\pm$ 0.11 & 0.07 $\pm$  0.16 & 0.87 $\pm$ 0.23 & 0.19 & -4.6\\
\enddata
\end{deluxetable*}



\end{document}